\documentclass{JHEP3} %{JHEP3}
\JHEPspecialurl{http://jhep.sissa.it/JOURNAL/JHEP3.tar.gz}
\usepackage{amsmath,amssymb,epsfig}

 \DeclareMathOperator{\im}{Im}

\makeatletter
\newcommand{\fmslash}[2][0mu]{%
  \mathchoice
    {\fmsl@sh\displaystyle{#1}{#2}}%
    {\fmsl@sh\textstyle{#1}{#2}}%
    {\fmsl@sh\scriptstyle{#1}{#2}}%
    {\fmsl@sh\scriptscriptstyle{#1}{#2}}}
\newcommand{\fmsl@sh}[3]{%
  \m@th\ooalign{$\hfil#1\mkern#2/\hfil$\crcr$#1#3$}}

\newcommand{\tr}{\hbox{tr}}

\title{\begin{center}
 \bf  \Large  Neutrino propagation in noncommutative spacetimes
 \end{center}}

\author{R. Horvat$^1$, A. Ilakovac$^2$, P. Schupp$^3$, J. Trampeti\'c$^{1,4}$ and
J. You$^1$ \\
1. Rudjer Bo\v skovi\' c Institute,
P.O.Box 180, HR-10002 Zagreb, Croatia \\
2. Faculty of Science, University of Zagreb, Bijeni\v{c}ka 32 Zagreb, Croatia\\
3. Center for Mathematics, Modeling and Computing,
Jacobs University Bremen, Campus Ring 1, 28759 Bremen, Germany\\
4. Max-Planck-Institut f\"ur Physik, (Werner-Heisenberg-Institut),
  	 F\"ohringer Ring 6, D-80805 M\"unchen, Germany\\
E-mail: \email{horvat@lei3.irb.hr}, \email{ailakov@rosalind.phy.hr},
\email{p.schupp@jacobs-university.de}, \email{josipt@rex.irb.hr }, \email{youjiangyang@gmail.com}}

\received{\today}               %%
\accepted{\today}               %% These are for published papers.

\abstract{One-loop $\theta$-exact quantum corrections to the neutrino
propagator are computed in noncommutative
$\rm U_{\star}(1)$ gauge-theory based on Seiberg-Witten maps.
Our closed form results show that the one-loop
correction contains a hard $1/\epsilon$ UV divergence,
as well as  a logarithmic IR-divergent term of
the type $\ln\sqrt{(\theta p)^2}$, thus considerably softening
the usual UV/IR mixing phenomenon.
We show that both of these problematic terms vanish
for a certain choice of the noncommutative parameter $\theta$ which
preserves unitarity.
We find non-perturbative modifications of the neutrino 
dispersion relations which are assymptotically independent
of the scale of noncommutativity in both the low and high energy 
limits and may allow superluminal propagation.
Finally, we demonstrate how the prodigious
freedom in Seiberg-Witten maps may be used to
affect neutrino propagation in a profound way.}

\begin{document}
%\maketitle

%%%%%%%%%%%%%%%%%%%%%%%%%%%%%%%%%%%%%%%%%%%%%%%%%%%%%%%%%%%%%%%%%%%%%%%%%%%%%%%%%%%%%%%%%

\section{Introduction}

%In modern theoretical physics is well known that quantum field theory works well in describing the pertinent physics at least to the LHC scale ($\sim 2\times 10^{-20}$ m). Theoretical speculations about the question {\it what happens below this scale?} involve search for mathematical foundations of {\it spacetime quantization} that could produce the correct modification of quantum field theory which would be valid at such length scale.

The  study of spacetime quantization has  originally been motivated by
major problems of physics at
extremely-high energies, in particular the problems of renormalization
and quantum gravity. Heisenberg-type spacetime uncertainty relations
can effectively lead to a replacement of the
continuum of points by finite size spacetime cells, thus providing a means
by which to tame UV divergences.
A branch of mathematics arising from these motivations has come
to be known as {\it noncommutative geometry}.
It is reasonable to expect that noncommutative~(NC) field theory models can
provide  some guidance for a deeper understanding of the structure of spacetime
at extremely-high energies. In fact, these NC models appear quite naturally
in string theory~\cite{Seiberg:1999vs}.
The relevant scale of noncommutativity may very well be beyond direct experimental
reach for the foreseeable future (except in certain theories with large extra dimensions).
Nevertheless, non-perturbative effects can nevertheless lead
to profound observable consequences for low energy physics.
A famous example is UV/IR mixing. Another striking example is
the running of the coupling constant of noncommutative U(1) gauge theory
in the simple {\it star($\star$)-product} formalism \cite{Martin:1999aq}.
The beta function
\begin{equation}\label{beta_function}
\beta(g^2) = -\frac{1}{8 \pi^2} \frac{22}{3} g^4 N^2
\end{equation}
of NC $\rm U(N)$ gauge theory is identical to that of ordinary $\rm SU(N)$
Yang-Mills theory for $N>1$, but in the noncommutative case
it remains valid even for Abelian $N=1$ gauge theory.
Hence the theory will suffer from asymptotic freedom
\cite{Martin:1999aq,Ettefaghi:2010zz}\footnote {The negative $\beta$-function is  most significant in pure
noncommutative gauge theory. When fermion fields are added the situation
can change considerably \cite{Ettefaghi:2010zz}.}.
This result is manifestly independent of the scale of noncommutativity
and thus remains valid even for vanishingly small (but non-zero) noncommutivity.

%In this article we derive non-perturbative result for neutrino propagation in noncommutative spacetimes, including in particular modified dispersion relations that allow superluminal velocities as they are currently discussed in the context of the OPERA experiment~\cite{:2011zb}.

In a simple model of NC spacetime local coordinates
$x^\mu$ are promoted to hermitian operators
$\hat x^\mu$ satisfying spacetime noncommutative relations
\begin{equation}
[\hat x^\mu ,\hat x^\nu]=i\theta^{\mu\nu},
\label{x*x}
\end{equation}
where $\theta^{\mu\nu}$ is real antisymmetric matrix of dimension $length^2$.
The commutator (\ref{x*x}) implies spacetime uncertainty relations
\begin{equation}
\Delta x^\mu \Delta x^\nu \geq \frac{1}{2}|\theta^{\mu\nu}|.
\label{deltax*x}
\end{equation}
%The introduced theoretical physics framework is called {\it noncommutative field theory} (NFT)
%and it may be a relevant effective model to describe physics at scales between
%the Planck scale
%and the LHC scale.
It is straight-forward to formulate field theories on such
noncommutative spaces as a deformation of the ordinary field theories.
The noncommutative deformation is implemented by replacing the usual pointwise
product of a pair of fields $f(x)$ and $g(x)$ by a $\star$-product  in the action:
\begin{equation}
f(x)g(x) \longrightarrow (f\star g)(x)=f(x)g(x)
+ {\cal O}(\theta,\partial f,\partial g).
\label{fg}
\end{equation}
The Moyal-Weyl $\star$-product is relevant for the case of
a constant $\theta^{\mu\nu}$ and is defined as follows:
\begin{equation}
(f\star g)(x)=f(x)
e^{\frac{i}{2}\overleftarrow{{\partial}_\mu}\,
\theta^{\mu\nu}\,\overrightarrow{{\partial}_\nu}} g(x).
\label{f*g}
\end{equation}
(The $\star$-product has also an alternative integral formulation,
making its non-local character more transparent.)
The operator commutation relation (\ref{x*x}) is then realized by the {\it star($\star$)-commutator}
\begin{equation}
[\hat x^\mu ,\hat x^\nu]=[x^\mu \stackrel{\star}{,} x^\nu]=i\theta^{\mu\nu}.
\end{equation}
%This introduces field operators ordering ambiguities
%and breaks ordinary gauge invariance,
%because local gauge transformations do not commute with star products.

In analogy to the introduction of covariant derivatives in gauge field theory,
the star-product can be promoted to a gauge-field dependent covariant star product.
Together with a gauge-field dependent covariant integral measure
this generally leads to a noncommutative gauge field theory based
on so-called Seiberg-Witten (SW) maps~\cite{Seiberg:1999vs} .
The resulting type of noncommutative quantum field theory has been
studied for quite some time.

In this construction the {\it noncommutative fields} are obtained via SW maps from
the original {\it commutative fields}. It is important to note that there is typically some freedom in the choice of SW map and that there is no warranty that every change in the choice of SW map will lead to a physically equivalent theory:
Deformation, like quantization, is usually not unique and different deformations can lead to physically inequivalent models. The deformed model is not uniquely fixed by its commutative classical (tree level) $\theta \to 0$ limit.
Different SW maps can behave like different quantization procedures.
%It is commutative instead of the noncommutative gauge symmetry that is
%preserved as the fundamental symmetry of the theory.
%and their generalizations
%Despite various progress in this branch, the exact loop computation still remains in general a
%challenging open question. In this article we present an $\theta$-exact evaluation
%of the one-loop neutral fermion self-energy.

The perturbative quantization of noncommutative field theories was first
proposed in a pioneering paper about fifteen years ago~\cite{Filk:1996dm}.
Since then considerable efforts have been devoted to this subject.
However, despite some significant progress like the
models in \cite{Grosse:2004yu,Magnen:2008pd,Meljanac:2011cs}, a complete understanding
of quantum loop corrections still remains in general a challenging open question.
This fact is particularly true for the models constructed
by using Seiberg-Witten  map expansion
since the map was for a long time expressed as an approximative expansion in
the noncommutative parameter $\theta$
\cite{Calmet:2001na,Behr:2002wx,Aschieri:2002mc,Schupp:2002up,Minkowski:2003jg}.
One loop quantum properties
\cite{Bichl:2001nf,Bichl:2001cq,Grimstrup:2002af,Banerjee:2001un,Martin:2002nr,Brandt:2003fx,Buric:2006wm,
Latas:2007eu,Buric:2007ix,Martin:2009sg,Martin:2009vg,Tamarit:2009iy,Buric:2010wd},
as well as the related phenomenology \cite{Ohl:2004tn,Melic:2005su,Tamarit:2008vy,
Alboteanu:2006hh,Alboteanu:2007by,Buric:2007qx}
of the $\theta$-expanded models, have also been investigated recently.

The tree and one-loop analysis of the minimal NCSM truncated to
first order in $\theta$ \cite{Buric:2006wm,Latas:2007eu,Buric:2007qx},
has shown that considerably
different physical behavior can arise from different choices of the Seiberg-Witten map. Furthermore,
an analysis of the photon two-point function in a $\theta$-expanded model up to all
orders via SW map \cite{Bichl:2001cq}\footnote{To absorb the loop divergences, at each order in $\theta$ the freedom in  the choice of the Seiberg-Witten map has been used
\cite{Bichl:2001cq,Grimstrup:2002af}.}
reveals that the parameters that fix the choice of SW map are running coupling constants \cite{Grimstrup:2002af}.
This again indicates physical differences among different SW map deformations.

Some results about closed-form solutions and/or alternative $\theta$-exact approaches,
starting from exact solutions for the Seiberg-Witten map, have existed for quite some
time~\cite{Mehen:2000vs,Jurco:2001my,Okawa:2001mv,Barnich:2003wq}.
Quite recently, $\theta$-exact SW map expansions,
in the framework of covariant noncommutative quantum gauge field theory
\cite{Jackiw:2001jb},
were applied in loop computation \cite{Zeiner:2007,Schupp:2008fs,Horvat:2011bs}
and phenomenology \cite{Horvat:2010sr,Horvat:2011iv}.
These more sophisticated theories differ quite drastically from their
$\theta$-expanded cousins, as they introduce in general a nonstandard
denominator $p\theta q$ into the loop integral. A few methods have
been proposed so far to handle this problem: An expansion and re-summation with respect
to $\theta$ in the loop integral allowed some progress in \cite{Zeiner:2007};
another approximative method which also looks promising is integration using
parametric derivatives. In general however, obtaining result in a closed form still remains a
challenging problem \cite{Schupp:2008fs}.

In this article we obtain  a closed formula for
the one-loop correction to the propagator of
a massless neutrino (neutral fermion) in
the adjoint representation of $\rm U_\star(1)$ gauge group
\cite{Horvat:2011bs}.
In the evaluation  we combine parameterizations of Schwinger, Feynman, and
a (modified) HQET parameterization \cite{Grozin:2000cm},
which was developed originally for heavy quark effective
theory (HQET). This model is relatively easy to handle since both
the gauge field and the fermion Seiberg-Witten map can be defined using
generalized star products. The model is interesting in its own right for both
theory and phenomenology: Its unexpanded version features  a fermion-boson
number symmetry which can cancel the leading order ultraviolet/infrared
(UV/IR) mixing \cite{Matusis:2000jf}.
The model was also considered as an example
for tree level neutrino-photon
coupling via noncommutativity \cite{Schupp:2002up, Minkowski:2003jg}.

The radiative corrections that we obtain contain in general both
a hard UV term and a logarithmic IR singularity. At a special value of
the noncommutative parameter $\theta$, both singularities vanishes.
%The usual vacuum polarization of photon from $\star$-product only
%NCQED shows some similar behavior \cite{Hayakawa:1999yt,Hayakawa:1999zf,
%Matusis:2000jf,Hayakawa:2000zi,Brandt:2001ud,Brandt:2002if}.
%Essential difference to \cite{Hayakawa:1999yt,Hayakawa:1999zf,Hayakawa:2000zi}
%is that in  our case both terms are proportional to the scale-independent structure of
%the NC dependent nonlocal factor, so called $\theta$-ratio.
Analyzing the poles of the resulting (finite) modified neutrino propagator
reveals, depending on the energy regime,  modes with either heavy masses or
modes whose propagation depends on the preferred direction in space set by
$\theta^{\mu\nu}$. In the low-energy regime we find modes propagating superluminaly.
%In particular, it depends on celebrated UV/IR mixing term and some of
%the propagating modes acquires mass.
%Clearly nonlocality of starting action turns into the nonlocality of
%the final results, which thus reflects their nonperturbative character. Besides closeness,
%our result also presents explicit differences between gauge theories
%with and without SW map expansion.
These properties present some previously unknown features of $\theta$-exact
noncommutative quantum field theories (NCQFT).

The article is structured as follows:
In the following section we describe the actions of two alternative models which differ with regard to the choice of SW map and we
give the relevant Feynman rules.  Section~3 is devoted to
the computation of the one-loop neutrino self-energy. Nonequivalent
divergences and corresponding dispersion relations
for both actions are described.
Asymptotic dispersion relations are given for
the low-energy and the high-energy regimes.
Section~4 is devoted to discussion and conclusions.
Relevant computational details of the nontrivial
loop-integrals are given in two appendixes.

\section{Model}

In setting up our models we adhear to the following principles of $\theta$-exact NCGFT:  \\
(i) The main principles that we are implementing in the construction of all of our
$\theta$-exact noncommutative models are: The standard field content and
the commutative gauge symmetry as the fundamental symmetry of the theory are fixed.\\
(ii) In the construction of the noncommutative action, generically,
electrically neutral matter fields will
be promoted via (hybrid) Seiberg-Witten maps to noncommutative fields that
couple to photons and transform in the adjoint representation of $\rm U_{\star}(1)$.\\
(iii) The different actions discussed in this paper are constructed by employing SW map freedom.\\
(iv) The inclusion of all gauge covariant coupling terms into these actions is
 a prerequisite for reasonable UV behavior.\\

\subsection{Action}

Taking the above into account we arrive at the following model of the
SW map type noncommutative $\rm U_{\star}(1)$ gauge theory with a gauge field $A_\mu$
coupled to a massless neutral fermion $\Psi$ via a star-commutator
\cite{Schupp:2002up,Minkowski:2003jg}
\begin{equation}
[A_\mu\stackrel{\star}{,}\Psi]=A_\mu\star \Psi-\Psi\star A_\mu.
\label{A-Psi}
\end{equation}
The action is defined in the usual way \cite{Schupp:2002up,Minkowski:2003jg,Schupp:2008fs}
\begin{equation}
S=\int-\frac{1}{4}F^{\mu\nu}\star F_{\mu\nu}+i\bar\Psi\star\fmslash{D}\Psi\,,
\label{S}
\end{equation}
with definitions of the noncommutative covariant derivative
and field strength resembling the corresponding expressions of non-abelian Yang-Mills theory:
\begin{gather*}
D_\mu\Psi=\partial_\mu\Psi-i[A_\mu\stackrel{\star}{,}\Psi]\quad\mbox{and}\quad
F_{\mu\nu}=\partial_\mu A_\nu-\partial_\nu
A_\mu-i[A_\mu\stackrel{\star}{,}A_\nu].
\label{DF}
\end{gather*}
All the NC fields in this action are images of the corresponding commutative fields
$a_\mu$ and $\psi$ under (hybrid) Seiberg-Witten maps. In the original work of
Seiberg and Witten and majority of the subsequent applications,
these maps are understood as power series of the noncommutativity parameter
$\theta^{\mu\nu}$. Physically, this corresponds to an expansion in momenta and is valid
only for low energy phenomena.
%Here we shall not subscribe to this point of view and instead interpret the noncommutative
%fields as valued in the enveloping algebra of the underlying gauge group.
Here we give an alternative point of view and employ an expansion in  formal powers of
the gauge field $a_\mu$ and hence in powers of
the coupling constant $e$. At each order in $a_\mu$ we shall determine $\theta$-exact expressions.
In the following we discuss the model construction
for the massless case, and we shall set $e=1$. To restore the coupling
constant one simply substitutes $a_\mu$ by $ea_\mu$ and then divides
the gauge-field term in the Lagrangian by $e^2$.

Next step we expand the action in terms of the commutative fields $a_\mu$ and $\psi$
using the following SW map solution \cite{Schupp:2008fs}
\begin{equation}
\begin{split}
A_\mu&=\,a_\mu-\frac{1}{2}\theta^{\nu\rho}{a_\nu}\star_2(\partial_\rho
a_\mu+f_{\rho\mu})+\mathcal O(a^3),
\\
\Psi&=\psi-\theta^{\mu\nu}
{a_\mu}\star_2{\partial_\nu}\psi+\frac{1}{2}\theta^{\mu\nu}\theta^{\rho\sigma}
\bigg\{(a_\rho\star_2(\partial_\sigma
a_\mu+f_{\sigma\mu}))\star_2{\partial_\nu}\psi+2a_\mu{\star_2}
(\partial_\nu(a_\rho{\star_2}\partial_\sigma\psi))\\&-
a_\mu{\star_2}(\partial_\rho
a_\nu{\star_2}\partial_\sigma\psi)-\big[a_\rho\partial_\mu\psi(\partial_\nu
a_\sigma+f_{\nu\sigma})-\partial_\rho\partial_\mu\psi a_\nu
a_\sigma\big]_{\star_3}\bigg\}+\mathcal O(a^3)\psi.
\end{split}
\label{SWmap}
\end{equation}
Here the two generalized star products
\begin{gather}
f(x)\star_2 g(x)=\frac{\sin\frac{\partial_1\theta
\partial_2}{2}}{\frac{\partial_1\theta
\partial_2}{2}}f(x_1)g(x_2)\bigg|_{x_1=x_2=x}\,,
\label{f*2g}\\
[f(x)g(x)h(x)]_{\star_3}=\bigg[\frac{\sin(\frac{\partial_2\theta
\partial_3}{2})\sin(\frac{\partial_1\theta(\partial_2+\partial_3)}{2})}
{\frac{(\partial_1+\partial_2)\theta \partial_3}{2}
\frac{\partial_1\theta(\partial_2+\partial_3)}{2}}
+\{1\leftrightarrow 2\}\,\bigg]f(x_1)g(x_2)h(x_3)\bigg|_{x_i=x}\,,
\label{fgh*3}
\end{gather}
are symmetric in their arguments, but non\-associative.
The resulting expansion in the coupling constant defines
the one-photon-two-fermion, two-photon-two-fermion and three-photon vertices
$\theta$-exactly.

The expansion of the action is straightforward using the SW map  \eqref{SWmap}.
The first nontrivial contribution in the gauge field expansion leads
to the following photon self-interaction terms
\begin{equation}
\begin{split}
S_g=&\int \;i\partial_\mu a_\nu\star[a^\mu\stackrel{\star}{,}a^\nu]+\frac{1}{2}\partial_\mu
\left(\theta^{\rho\sigma}a_\rho\star_2(\partial_\sigma a_{\nu}+f_{\sigma\nu})\right)\star
f^{\mu\nu}+\mathcal O(a^4).
\end{split}
\label{Sgauge}
\end{equation}
The photon-fermion interaction up to 2-photon-2-fermion terms is derived
using the first order gauge field and second order fermion field expansion
\begin{equation}
\begin{split}
S_f&=\int \;\bar\psi\gamma^\mu[a_\mu\stackrel{\star}{,}\psi]
+i(\theta^{ij}\partial_i\bar\psi
\star_2 a_j)\fmslash\partial\psi-i\bar\psi\star
\fmslash\partial(\theta^{ij}
a_i\star_2\partial_j\psi)+(\theta^{ij}\partial_i\bar\psi \star_2
a_j)\gamma^\mu[a_\mu\stackrel{\star}{,}\psi]
\\&
-\!\bar\psi\gamma^\mu[a_\mu\!\stackrel{\star}{,}\!\theta^{ij}
a_i\!\star_2\!\partial_j\psi]\!-\!\frac{1}{2}\bar\psi\gamma^\mu
[\theta^{ij}a_i\!\star_2\!(\partial_j
a_\mu\!+\!f_{j\mu})\!\stackrel{\star}{,}\!\psi]\!
-\!i(\theta^{ij}\partial_i\bar\psi
\!\star_2\!a_j)\fmslash\partial(\theta^{kl}
a_k\!\star_2\!\partial_l\psi)
\\&
+\frac{i}{2}\theta^{ij}\theta^{kl}
\bigg((a_k\star_2(\partial_l
a_i+f_{li}))\star_2\partial_j\bar\psi
+2a_i\star_2(\partial_j(a_k\star_2\partial_l\bar\psi))-
a_i\star_2(\partial_k
a_j\star_2\partial_l\bar\psi)
\\&
+\big[a_i\partial_k\bar\psi(\partial_j
a_l+f_{jl})-\partial_k\partial_i\bar\psi a_j a_l\big]_{\star_3}\bigg)
\fmslash\partial\psi+\frac{i}{2}\theta^{ij}\theta^{kl}\bar\psi\fmslash\partial
\bigg((a_k\star_2(\partial_l
a_i+f_{li}))\star_2\partial_j\psi
\\&
+\!2a_i\!\star_2\!
(\partial_j(a_k\!\star_2\!\partial_l\psi))\!-\!a_i\!\star_2\!(\partial_k
a_j\!\star_2\!\partial_l\psi)\!+
%\!\frac{1}{2}\theta^{ij}\theta^{kl}
\big[a_i\partial_k\psi(\partial_j
a_l\!+\!f_{jl})\!-\!\partial_k\partial_i\psi a_j
a_l\big]_{\star_3}\bigg)
\\&
+\bar\psi\mathcal{O}(a^3)\psi.
\end{split}
\label{Sfermion}
\end{equation}
It is important to note that the actions (\ref{Sgauge}) and (\ref{Sfermion})
for gauge and matter fields  respectively, are nonlocal
objects due to the presence of the non-local (generalized) star
products~$\star$, $\star_2$ and $\star_3$.
The appearance of non-locality in the actions is expected to be reflected in corresponding
quantum corrections to the neutrino propagator (2-point functions).

\subsection{Feynman rules}

From the interaction terms we read out the three vertices needed for
one-loop two point-function computations \cite{Horvat:2011qn}
\begin{figure}
\begin{center}
\includegraphics[width=12cm,height=5cm]{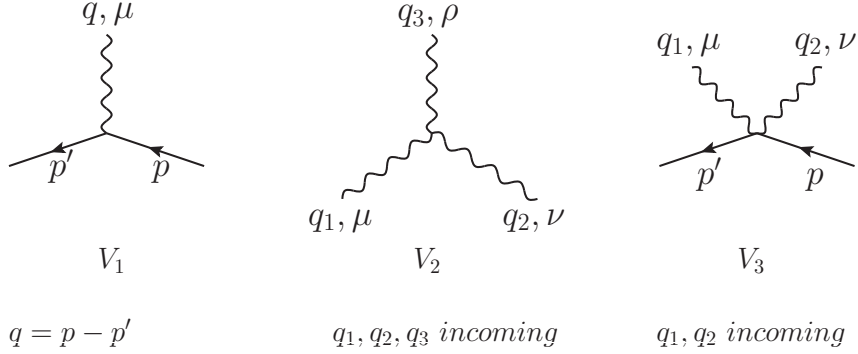}
\caption{Three- and four-field vertices}
\end{center}
\label{fig:3-4}
\end{figure}
\begin{eqnarray}
V_1^\mu(q,p)
 &=&
-F(q,p)[(q\theta p)\gamma^\mu+\fmslash{p}\tilde{q}^\mu-\fmslash{q}\tilde{p}^\mu]
%\frac{1\pm \gamma_5}{2},
,\;\;\;F(p,q)=\frac{\sin\frac{q\theta p}{2}}{\frac{q\theta p}{2}}\,,
\label{V1}\\
\lefteqn{V_2^{\mu\nu\rho}(q_1,q_2,q_3)
 =
 - 2 \bigg\{\sin\frac{q_1\theta q_2}{2}
 [ (q_1-q_2)^{\rho} g^{\mu\nu}
 + (q_2-q_3)^{\mu} g^{\nu\rho}
 + (q_3-q_1)^{\nu} g^{\rho\mu}]}
\nonumber\\
 &+&
    F(q_1,q_2)
 \Big[
   \theta^{\mu\nu} (q_2q_3 q_1^{\rho} - q_1q_3 q_2^{\rho})
 + \theta^{\mu\rho} (q_2q_3 q_1^{\nu} - q_1q_2 q_3^{\nu})
%\nonumber\\
% &&
 + \theta^{\nu\rho} (q_1q_3 q_2^{\mu} - q_1q_2 q_3^{\mu})
\nonumber\\
 &-&
   g^{\mu\nu}(q_2^2\tilde{q}_1^{\rho} + q_1^2 \tilde{q}_2^{\rho})
 - g^{\mu\rho}(q_1^2\tilde{q}_3^{\nu} + q_3^2 \tilde{q}_1^{\nu})
 - g^{\nu\rho}(q_3^2\tilde{q}_2^{\mu} + q_2^2 \tilde{q}_3^{\mu})
\nonumber\\
 &+&
   q_3^{\rho}(\tilde{q}_2^{\mu}q_3^{\nu} + \tilde{q}_1^{\nu}q_3^{\mu})
 + q_2^{\nu}(\tilde{q}_1^{\rho}q_2^{\mu} + \tilde{q}_3^{\mu}q_2^{\rho})
 + q_1^{\mu}(\tilde{q}_2^{\rho}q_1^{\nu} + \tilde{q}_3^{\nu}q_1^{\rho})\Big]\bigg\}\,,
\label{V2}
\end{eqnarray}
\begin{eqnarray}
\lefteqn{V^{\mu\nu}_{3}(q_1,q_2,p,p')
=2i\bigg\{2\frac{\sin\frac{q_1\theta p}{2}\sin\frac{q_2\theta p'}{2}}
{q_1\theta p}\tilde p^{\mu}\gamma^{\nu}-
2i\frac{\sin\frac{q_1\theta p}{2}\sin\frac{q_2\theta p'}{2}}
{q_2\theta p'}\tilde {p}^{'\nu}\gamma^{\mu}}
\nonumber\\
&-&
 \frac{\sin\frac{p\theta p'}{2}\sin\frac{q_1\theta q_2}{2}}{q_1\theta q_2}
 (2\gamma^{\nu}\tilde q_2^{\mu}-\fmslash q_2\theta^{\mu\nu})
-4i\frac{\sin\frac{q_1\theta p}{2}\sin\frac{q_2\theta p'}{2}}{q_1\theta p q_2\theta p'}
 (\fmslash q_2+\fmslash p')\tilde p^{\mu}\tilde p^{'\nu}
\nonumber\\
&+&
 \fmslash p'\bigg[\frac{\sin\frac{p\theta p'}{2}
 \sin\frac{q_1\theta q_2}{2}}{q_1\theta q_2 p\theta p'}
 (q_2\theta p\theta^{\mu\nu}
-2\tilde q_2^{\mu}\tilde p^{\nu})
\nonumber\\
&-&
 \frac{\sin\frac{q_1\theta p'}{2}\sin\frac{q_2\theta p}{2}}{p_1\theta p' q_2\theta p}
 2(\tilde q_2-\tilde p)^{\mu}\tilde p^{\nu}
+\frac{\sin\frac{q_1\theta p'}{2}\sin\frac{q_2\theta p}{2}}{q_1\theta p}\theta^{\mu\nu}
\nonumber\\
&+&
 \bigg(\frac{\sin\frac{q_2\theta p}{2}\sin\frac{q_1\theta p'}{2}}{q_2\theta p' q_1\theta p'}
 +\frac{\sin\frac{q_1\theta q_2}{2}\sin\frac{p\theta  p'}{2}}{q_2\theta p' p\theta p'}\bigg)
 (2\tilde p^{\nu}\tilde q_2^{\mu}+\theta^{\mu\nu}p\theta q_2
 -\tilde p^{\mu}\tilde p^{\nu})\bigg]
\nonumber\\
&+&
 \fmslash p\bigg[\frac{\sin\frac{p'\theta p}{2}
 \sin\frac{q_1\theta q_2}{2}}{q_1\theta q_2 p'\theta p}
 (2\tilde q_2^{\mu}\tilde p^{'\nu}-q_2\theta p'\theta^{\mu\nu})
\nonumber\\
&+&
 \frac{\sin\frac{q_1\theta p}{2}\sin\frac{q_2\theta p'}{2}}{q_1\theta p q_2\theta p'}
 2(\tilde q_2+\tilde p')^{\mu}\tilde p^{'\nu}
-\frac{\sin\frac{q_1\theta p}{2}\sin\frac{q_2\theta p'}{2}}{q_1\theta p}\theta^{\mu\nu}
\nonumber\\
&-&
\bigg(\frac{\sin\frac{q_2\theta p'}{2}\sin\frac{q_1\theta p}{2}}{q_2\theta p p_1\theta p}
+\frac{\sin\frac{q_2\theta q_1}{2}\sin\frac{p'\theta p}{2}}{q_2\theta p p'\theta p}\bigg)
(2\tilde p^{'\nu}\tilde q_2^{\mu}+\theta^{\mu\nu}p'\theta q_2
+\tilde p^{'\mu}\tilde p^{'\nu})\bigg]\bigg\}
%\frac{1\pm \gamma_5}{2}
\nonumber\\
&+&
\{q_1\leftrightarrow q_2\,\mbox{and}\,
\mu\leftrightarrow \nu\}\,,
\label{V3}
\end{eqnarray}
where ${\tilde q}^\mu=(\theta q)^{\mu}=\theta^{\mu\nu} q_{\nu}$, and in addition we define
${\tilde{\tilde q}}^\mu=(\theta\theta q)^\mu=\theta^{\mu\nu}\theta_{\nu\rho} q^{\rho}$\,.

\subsection{Alternative actions and Feynman rules}

As discussed in \cite{Horvat:2011iv}, there exist alternative consistent
and covariant choices for the noncommutative  interactions.
This is related to a  freedom in the choice of SW maps used in
the construction of the theory.
For the action \eqref{Sfermion} and Feynman rules (\ref{V1}), (\ref{V3}) ,
we presented in this section an alternative construction
of the massless action, with coupling constant $e=1$.

We start with the action for a neutral massless free fermion field
\begin{equation}
S_f = \int \bar\psi\gamma^\mu\partial_\mu\psi \,
=\int  \bar\psi\star \gamma^\mu\partial_\mu\psi\, \,,
\label{A1}
\end{equation}
where, as indicated, a Moyal-Weyl type star product can be inserted
or removed by partial integration.
Following the method of constructing a covariant NC gauge theory outlined
at the beginning of this section,
we lift the factors in the action via (generalized) Seiberg-Witten maps
$\hat \Psi[\bar\psi,...]$, $\hat \Phi[\partial_\mu\psi,...]$
to noncommutative status as follows:
\begin{equation}
S_f \to S_{f_{alt}}=\int  \hat\Psi(\bar\psi)\gamma^\mu \hat\Phi(\partial_\mu\psi) \,
= \int  \hat\Psi(\bar\psi)\star \gamma^\mu \hat\Phi(\partial_\mu\psi)  \,  \,.
\label{A2}
\end{equation}
Now if the SW maps $\hat\Psi$, $\hat\Phi$ and a corresponding map $\hat\Lambda$
for the gauge parameter $\lambda$ satisfy
\begin{equation}
\begin{split}
\delta_\lambda(\hat\Psi(\bar\psi))=
i[\hat\Lambda(\lambda)\stackrel{\star}{,}\hat\Psi(\bar\psi)],\\
\delta_\lambda(\hat\Phi(\partial_\mu\psi))
=i[\hat\Lambda(\lambda)\stackrel{\star}{,}\hat\Phi(\partial_\mu\psi)]\,,
\end{split}
\label{A3}
\end{equation}
we will have a noncommutative action which is
gauge invariant under infinitesimal commutative gauge
transformations $\delta_\lambda$ and reduces to the free fermion action
in the commutative limit $\theta\to 0$.

The appropriate map $\hat\Psi$ can be the same as in \eqref{SWmap}.
Recalling that we are dealing with neutral fields, i.e.
$\delta \psi = 0$ and $\delta (\partial_\mu\psi) = 0$,
we notice that we can in principle use the same map also for
$\hat\Phi$:
\begin{equation}
\begin{split}
\hat\Phi_{alt2}(\partial_\mu\psi)&=\hat\Psi(\partial_\mu\psi)=\Psi(\psi \to \partial_\mu\psi)
=\partial_\mu\psi-\theta^{ij}
{a_i}\star_2\partial_j(\partial_\mu\psi)
\\&+\frac{1}{2}\theta^{ij}\theta^{kl}
\bigg\{(a_k\star_2(\partial_l
a_i+f_{li}))\star_2{\partial_j}(\partial_\mu\psi)+2a_i{\star_2}
(\partial_j(a_k{\star_2}\partial_l(\partial_\mu\psi)))
\\&-
a_i{\star_2}(\partial_k
a_j{\star_2}\partial_l(\partial_\mu\psi))-\big[a_k\partial_i(\partial_\mu\psi)(\partial_j
a_l+f_{jl})-\partial_k\partial_i(\partial_\mu\psi) a_j
a_l\big]_{\star_3}\bigg\}+\mathcal O(a^3)\psi \,.
\end{split}
\label{A5}
\end{equation}
This construction is quite unusual from the point of gauge theory,
as it yields a covariant derivative
term without introducing a covariant derivative. In any case the resulting action
\begin{equation}
\begin{split}
S_{f_{alt2}}=&\int \Big(i\bar\psi\fmslash\partial\psi-i\left(\theta^{ij}\partial_j\bar\psi
\star_2 a_i\right)\fmslash\partial\psi+i\bar\psi\left(\theta^{ij}
a_i\star_2\fmslash\partial\partial_j\psi\right) \Big) \, \;
\\
&+i\left(\theta^{ij}\partial_j\bar\psi
\star_2 a_i\right)(\theta^{kl}
{a_k}\star_2\partial_l(\fmslash\partial\psi))
\\
&-\frac{i}{2}\theta^{ij}\theta^{kl}\bigg\{(a_k\star_2(\partial_l
a_i+f_{li}))\star_2{\partial_j}\bar\psi+2a_i{\star_2}
(\partial_j(a_k{\star_2}\partial_l\bar\psi))
\\
&-a_i{\star_2}(\partial_k
a_j{\star_2}\bar\psi)-\big[a_k\partial_i\bar\psi(\partial_j
a_l+f_{jl})-\partial_k\partial_i(\bar\psi) a_j
a_l\big]_{\star_3}\bigg\}\fmslash\partial\psi
\\
&+\frac{i}{2}\theta^{ij}\theta^{kl}\bar\psi
\bigg\{(a_k\star_2(\partial_l
a_i+f_{li}))\star_2{\partial_j}(\fmslash\partial\psi)+2a_i{\star_2}
(\partial_j(a_k{\star_2}\partial_l(\fmslash\partial\psi)))
\\
&-a_i{\star_2}(\partial_k
a_j{\star_2}\partial_l(\fmslash\partial\psi))-\big[a_k\partial_i(\fmslash\partial\psi)(\partial_j
a_l+f_{jl})-\partial_k\partial_i(\fmslash\partial\psi) a_j
a_l\big]_{\star_3}\bigg\}
\\
&+ \mathcal{O}(a^3)\,,
\end{split}
\label{A6}
\end{equation}
is consistent and gauge invariant.
The action leads to the following photon-fermion interaction vertices, i.e. Feynman rule,
%\begin{gather}
%\begin{split}
\begin{eqnarray}
&&V^{\mu}_{1_{alt2}}(q,p)= -F(q,p) (\theta q)^\mu\fmslash p
%\frac{1\pm \gamma_5}{2}
\,,
\label{V2alt2}\\
&&V^{\mu\nu}_{3_{alt2}}(q_1,q_2,p,p')=2i\fmslash p'\bigg\{-2
\frac{\sin\frac{q_1\theta p}{2}\sin\frac{q_2\theta p'}{2}}{q_1\theta p q_2\theta p'}
 \tilde p^{\mu}\tilde p^{'\nu}
\nonumber\\
&&+\bigg[
\frac{\sin\frac{p\theta p'}{2}
 \sin\frac{q_1\theta q_2}{2}}{q_1\theta q_2 p\theta p'}
 (q_2\theta q_1\theta^{\mu\nu}
-2\tilde q_2^{\mu}\tilde p^{\nu}+2\tilde q_2^{\mu}\tilde p^{'\nu})
\nonumber\\
&&-\frac{\sin\frac{q_1\theta p'}{2}\sin\frac{q_2\theta p}{2}}{q_1\theta p' q_2\theta p}
 2(\tilde q_2-\tilde p)^{\mu}\tilde p^{\nu}
+\bigg(\frac{\sin\frac{q_1\theta p'}{2}\sin\frac{q_2\theta p}{2}}{q_1\theta p'}
-\frac{\sin\frac{q_1\theta p}{2}\sin\frac{q_2\theta p'}{2}}{q_1\theta p}\bigg)\theta^{\mu\nu}
\nonumber\\
&&+\bigg(\frac{\sin\frac{q_2\theta p}{2}\sin\frac{q_1\theta p'}{2}}{q_2\theta p' q_1\theta p'}
+\frac{\sin\frac{q_1\theta q_2}{2}\sin\frac{p\theta  p'}{2}}{q_2\theta p' p\theta p'}\bigg)
(2\tilde p^{\nu}\tilde q_2^{\mu}+\theta^{\mu\nu}p\theta q_2-\tilde p^{\mu}\tilde p^{\nu})
\nonumber\\
&&+\frac{\sin\frac{q_1\theta p}{2}\sin\frac{q_2\theta p'}{2}}{q_1\theta p q_2\theta p'}
 2(\tilde q_2+\tilde p')^{\mu}\tilde p'^{\nu}
-\bigg(\frac{\sin\frac{q_2\theta p'}{2}\sin\frac{q_1\theta p}{2}}{q_2\theta p q_1\theta p}
+\frac{\sin\frac{q_2\theta p_1}{2}\sin\frac{p'\theta p}{2}}{q_2\theta p p'\theta p}\bigg)
\nonumber\\
&&\cdot(2\tilde p^{'\nu}\tilde q_2^{\mu}+\theta^{\mu\nu}p'\theta q_2
+\tilde p^{'\mu}\tilde p^{'\nu})\bigg]
\bigg\}
%\frac{1\pm \gamma_5}{2}
+\{q_1\leftrightarrow q_2\,\mbox{and}\,
\mu\leftrightarrow \nu\}\,.
%\end{split}
\label{V3alt2}
%\end{gather}
\end{eqnarray}
%\subsection{Action 3}

There is also a second choice for $\hat\Phi$:
\begin{equation}
\hat\Phi_{alt3}(\partial_\mu\psi)=D_\mu \hat\Psi(\psi) \,.
\label{A10}
\end{equation}
%This second choice of SW map
%differs from the first one by the gauge invariant term
%$\theta^{ij}f_{i\mu}\star_2\partial_j\psi$,
%indicating a freedom in the choice of Seiberg-Witten map.
This leads back to the original action  discussed in the previous subsection.
In general one can chose any superposition of the two SW maps $\hat\Phi_{alt2}$
and $\hat\Phi_{alt3}$ indicating a freedom in the choice of Seiberg-Witten map.
Note that Feynman rule (\ref{V1})
is more natural from the point of view of gauge theory.
In this article we analyze two different actions (Feynman rules),
i.e. the original action  and its alternative  presented in this section.
These essentially different actions (\ref{Sfermion}) and (\ref{A6}),
despite having the same field content and gauge symmetry,
will be shown to produce different neutrino self-energies at one-loop in the next section.

%\begin{equation}
%\begin{split}
%\hat\Phi_{alt3}(\partial_\mu\psi)=&D^\star_\mu \hat\Psi(\psi)=\partial_\mu
%\hat\Psi(\psi)-i[A_\mu\stackrel{\star}{,}\hat\Psi(\psi)]\\
%=&\partial_\mu\psi-\theta^{ij}a_i\star_2\partial_j\partial_\mu\psi
%+\theta^{ij}f_{i\mu}\star_2\partial_j\psi
%+O(a^2)\psi \,,
%\end{split}
%\label{A10}
%\end{equation}
%based on the well-known NC QED-type covariant derivative. This second choice of SW map
%differs from the first one by the gauge invariant term
%$\theta^{ij}f_{i\mu}\star_2\partial_j\psi$, indicating
%a freedom in the choice of Seiberg-Witten map.
%This second choice leads again to the vertex (\ref{V1}), that is to
%\begin{equation}
%\Gamma_{alt3}^{\mu}=iF(q,p)\left[ (q\theta p)\gamma^{\mu} + \fmslash p (\theta q)^{\mu}
%-\fmslash q(\theta p)^{\mu}\right]\,.
%\label{A11}
%\end{equation}
%In general one can chose any superposition of the two SW maps $\hat\Phi_{alt2}$
%and $\hat\Phi_{alt3}$,
%but in this article we shall focus on the second choice (\ref{A11}) or (\ref{V1})
%as it is more natural from the point of view of gauge theory.

%Note that we have kept the explicit left/right
%polarization projectors for fermions dependences in all above Feynman rules.

Finally, the pure gauge field
action $S_g$ in the alternative action, remains the same as in the original action.
We keep $S_g$ and $V_2^{\mu\nu\rho}$
intact, that is we apply (\ref{V2}) in the computation of
the two-point function from the alternative action.

\section{One-loop neutrino self-energy}

\subsection{Diagrams}

According to the vertices defined in the last section, there are four diagrams
which contribute to the neutrino self-energy at one-loop
\begin{figure}
\begin{center}
\includegraphics[width=11cm,height=7cm]{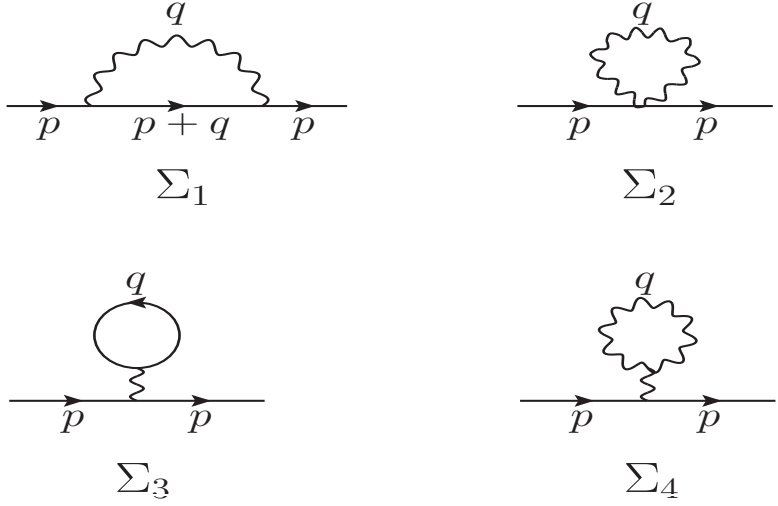}
\caption{One-loop contributions to neutrino self-energy}
\end{center}
\label{fig:1-loop}
\end{figure}
\begin{equation}
\Sigma_{1-loop}=\Sigma_1+\Sigma_2+\Sigma_3+\Sigma_4\,
\label{Sigma1-loop}
\end{equation}
The first $\Sigma_1$ and third $\Sigma_3$ are the analog of the commutative fermion self-energy.
The second one is similar to a tadpole graph in  $\phi^4$ theory. However,
it is not difficult to check, through direct computation from (\ref{V3}), that
%\begin{equation}
$g_{\mu\nu}V^{\mu\nu}_{3}(p,-p,q,q)=0$.
%\label{V3=0}
%\end{equation}
Therefore $\Sigma_2$ is equal to zero. The last two diagrams, $\Sigma_3$ and $\Sigma_4$
are tadpole diagrams obtained by connecting a one-loop one-point function to a triple
particle vertex. These two diagrams could be ruled out by the noncommutative
charge conjugation symmetry defined in \cite{Aschieri:2002mc},
so we do not take them into account.
Namely, here we have taken the charge conjugation transformation to be the same
as in the equations (64) to (66) from \cite{Aschieri:2002mc},
i.e. ${\theta^C}^{\mu\nu}=-\theta^{\mu\nu}$.
Thus only the two-point function bubble diagram $\Sigma_1$ needs to be evaluated.

\subsection{Two-point function}

The integral $\Sigma_1$ is defined as follows:
\begin{eqnarray}
\Sigma_1
&=&-\mu^{4-D}\int \frac{d^D
q}{(2\pi)^D}\frac{-ig_{\alpha\beta}}{q^2}\cdot
F(q,p)\left[(q\theta p)\gamma^\alpha+\fmslash
p\tilde q^\alpha-\fmslash q \tilde p^\alpha\right]\frac{i(\fmslash
p+\fmslash q)}{(p+q)^2}
\nonumber\\
&\cdot&  F(q,p)\left[-(q\theta
p)\gamma^\beta-\fmslash p\tilde q^\beta+\fmslash q \tilde
p^\beta\right]
%\frac{-1\mp\gamma^5}{2}
\nonumber\\
&=&-\mu^{4-D}\int \frac{d^D q}{(2\pi)^D}\bigg(\frac{\sin\frac{q\theta p}{2}}{\frac{q\theta
p}{2}}\bigg)^2\frac{1}{q^2}\frac{1}{(p+q)^2}
%\frac{1\pm\gamma^5}{2}
\bigg\{
(q\theta p)^2(4-D)(\fmslash p+\fmslash q)
\label{xi}\\
&+&(q\theta p)\left[/\!\!\tilde q(2p^2+2p\cdot q)-/\!\!\tilde
p(2q^2+2p\cdot q)\right]
\nonumber
\\
&+&\left[\fmslash p(\tilde q^2(p^2+2p\cdot q)-q^2(\tilde p^2+2\tilde
p\cdot\tilde q) )+\fmslash q(\tilde p^2(q^2+2p\cdot q)-p^2(\tilde
q^2+2\tilde p\cdot\tilde q ))\right]\bigg\}\,.
\nonumber
%\fmslash\xi\,,
\label{Sig1}
\end{eqnarray}
where $\mu$ represents a dimensionful regularization parameter.
Note that during the evaluation of $\Sigma _1$ we shall hide the
explicit $e$-dependence in the equations. We shall later make the
 $e$-dependence explicit  during the derivation
of the dispersion relations.

Combining the power of the $(q\theta p)$ term and the phase factor from
$F^2$, we can separate $\Sigma_1$ into a sum of three  integrals
\begin{equation}
-\Sigma_1=\left(4-D\right)I_{123}
+I_{456}+I_{789}\,,
\label{4D123}
\end{equation}
%where
\begin{eqnarray}
I_{123}&=&\int {\hspace{-2mm}}\frac{d^D
q}{(2\pi)^D}\frac{1}{q^2(p+q)^2}(\fmslash p+\fmslash q)
\left(2-e^{i q\theta p}-e^{-i q\theta p}\right),
\label{123}\\
I_{456}&=&\int{\hspace{-2mm}} \frac{d^D
q}{(2\pi)^D}\frac{1}{q^2(p+q)^2}\frac{1}{q\theta p}{\hspace{-1mm}}
\left[/\!\!\!\tilde q(2p^2+2p\cdot q)-/\!\!\!\tilde p(2q^2+2p\cdot
q)\right]{\hspace{-2mm}}\left(2-e^{i q\theta p}-e^{-i q\theta p}\right){\hspace{-1mm}},
\label{456}\\
I_{789}&=&\int{\hspace{-2mm}} \frac{d^D
q}{(2\pi)^D}\frac{1}{q^2(p+q)^2}\frac{1}{(q\theta
p)^2}\bigg[\fmslash p(\tilde q^2(p^2+2p\cdot q)-q^2(\tilde
p^2+2\tilde p\cdot\tilde q) )
\nonumber\\
&+&\fmslash q(\tilde p^2(q^2+2p\cdot
q)-p^2(\tilde q^2+2\tilde p\cdot\tilde q ))\bigg]
\left(2-e^{i q\theta p}-e^{-i q\theta p}\right),
\label{789}
\end{eqnarray}
where the presence of the regularization mass parameter $\mu$
in the above integrals is understood.
The explicit evaluation of the above loop-integrals, as given in  Appendix A, yields
%\subsection{Final expression for $\Sigma_1$ from actions 1 and/or 3}
%
%Finally with help of (\ref{B10})
%we can sum over the integrals \eqref{123}, \eqref{456} and \eqref{789} to obtain
\begin{equation}
\begin{split}
\Sigma_1&=-\frac{1}{(4\pi)^{\frac{D}{2}}}2/\!\!\! p p^2\bigg(\frac{\tr\theta\theta}{\tilde
p^2}+\frac{2\tilde{\tilde
p}^2}{\tilde p^4}\bigg)\bigg[(p^2)^{\frac{D}{2}-2}\Gamma\left(2-\frac{D}{2}\right)B
\left(\frac{D}{2}-1,\frac{D}{2}\right)
\\&-2\int\limits_0^1 dx(1-x)\left(x(1-x)p^2\right)^{\frac{D}{4}-1}
\,2^{\frac{D}{2}-2}(\tilde p^2)^{1-\frac{D}{4}}K_{2-\frac{D}{2}}
\left[\left(x(1-x)p^2\tilde
p^2\right)^{\frac{1}{2}}\right]\bigg]
\\
&-\frac{1}{(4\pi)^{\frac{D}{2}}}\bigg\{2\left(/\!\!\!
p\left(1-\frac{D}{2}\right)+\frac{p^2/\!\!\!\tilde{\tilde
p}}{\tilde p^2}-\frac{\tr\theta\theta}{2}\frac{p^2/\!\!\!
p}{\tilde p^2}\right)-\frac{/\!\!\!
p}{\tilde p^4}\left(\tilde{\tilde p}^2p^2-\tilde p^4\right)\bigg\}
\\
&\cdot\frac{\pi}{2\sin\frac{D\pi}{2}}
\int\limits_0^1 dx(1-x)(\tilde p^2)^{2-\frac{D}{2}}
\\
&\cdot\bigg[\left(x(1-x)p^2\tilde p^2\right)^{\frac{D}{2}-1}
\Gamma\left(\frac{1}{2}\right){}_1{\tilde F}_2\left(\frac{1}{2};
\frac{3}{2},\frac{D}{2};\frac{x(1-x)p^2\tilde p^2}{4}\right)
\\
&-2^{D-2}\Gamma\left(\frac{3-D}{2}\right){}_1{\tilde F}_2
\left(\frac{3-D}{2};\frac{4-D}{2},\frac{5-D}{2};
\frac{x(1-x)p^2\tilde p^2}{4}\right)\bigg].
\end{split}
\label{sigma1final}
\end{equation}

\subsection{Neutrino self-energy}

The general expression for the neutrino self-energy,
after evaluating integrals in (\ref{sigma1final}) for
$D=4-\epsilon$ and in $\epsilon\to 0$ limit,
receives the following closed form structure
\begin{equation}
\Sigma_1=-\gamma_{\mu}
\bigg[ p^{\mu}\: A + (\theta{\theta p})^{\mu}\; \frac{p^2}{(\theta p)^2}\;B\bigg]\,,
\label{sigma1AB}
\end{equation}
with
\begin{eqnarray}
A&=&\frac{1}{(4\pi)^2}\bigg[(s_1+2s_2)A_1 + (1+s_1+s_2)A_2\bigg]\,
\label{sigma1A12}\\
&=&
\frac{1}{(4\pi)^2}\;(s_1+2s_2)\;\bigg[\frac{2}{\epsilon}
+\ln \frac{\mu^2(\theta p)^2}{16}
+2-\psi_0\left(\frac{3}{2}\right)+\ln (4\pi)
\nonumber\\
&+&\frac{1}{2}\sum\limits_{k=1}^\infty\frac{\pi^{\frac{1}{2}}}{\Gamma\left(k+1\right)
\Gamma\left(k+\frac{3}{2}\right)}\left(\frac{p^2(\theta p)^2}{16}\right)^k
\left(\ln\frac{p^2(\theta p)^2}{16}-\psi_0\left(k+1\right)-\psi_0
\left(k+\frac{3}{2}\right)\right)\bigg]
\nonumber\\
&+&\frac{1}{(4\pi)^2}(1+s_1+s_2)
\bigg[2-\frac{1}{2}\sum\limits_{k=0}^\infty\frac{\pi^{\frac{1}{2}}
\Gamma\left(k+\frac{1}{2}\right)}{\Gamma\left(k+1\right)
\Gamma\left(k+\frac{3}{2}\right)\Gamma\left(k+\frac{5}{2}\right)}
\left(\frac{p^2(\theta p)^2}{16}\right)^{k+1}
\bigg(\ln\frac{p^2(\theta p)^2}{16}
\nonumber\\
&-&\psi_0\left(k+1\right)
+\psi_0\left(k+\frac{1}{2}\right)
-\psi_0\left(k+\frac{3}{2}\right)-\psi_0\left(k+\frac{5}{2}\right)\bigg)\bigg]\,,
\label{sigma1A}\\
%&=& %\end{equation}
%\begin{split}
B&=&\frac{-1}{(4\pi)^2}\;%\frac{p^2}{(\theta p)^2}\cdot
\bigg[4-\sum\limits_{k=0}^\infty\frac{\pi^{\frac{1}{2}}
\Gamma\left(k+\frac{1}{2}\right)}{\Gamma\left(k+1\right)
\Gamma\left(k+\frac{3}{2}\right)\Gamma\left(k+\frac{5}{2}\right)}
\left(\frac{p^2(\theta p)^2}{16}\right)^{k+1}
\bigg(\ln\frac{p^2(\theta p)^2}{16}
\nonumber\\
&-&\psi_0\left(k+1\right)
+\psi_0\left(k+\frac{1}{2}\right)%-\psi_0\left(k+1\right)
-\psi_0\left(k+\frac{3}{2}\right)-\psi_0\left(k+\frac{5}{2}\right)\bigg)\bigg]
=-\frac{2}{(4\pi)^2}A_2\,.
%\end{split}
\label{sigma1B}
\end{eqnarray}
Here $s_{1,2}$ are scale-independent $\theta$-ratios
\begin{equation}
s_1= p^2\frac{\tr\theta\theta}{(\theta p)^2},\;\;\;
s_2=p^2\frac{(\theta\theta p)^2}{(\theta p)^4},
\label{s1s2}
\end{equation}
and $A_{1,2}$ correspond to the first and
second square brackets in Eq. (\ref{sigma1A}), respectively.

It is important to note here that amongst other
terms contained in both coefficients $A_1$ and $A_2$, there are structures
proportional to
\begin{equation}
\left({p^2(\theta p)^2}\right)^{n+1}(\ln{(p^2(\theta p)^2)})^m,\;\;
{\forall n} \;\; {\rm and} \;\;m=0,1.
\label{lnUV/IR'}
\end{equation}
The numerical factors in front of the above structures are rapidly-decaying,
thus the series are always convergent for finite arguments
as we numerically demonstrated below:
\begin{eqnarray}
A_1&\simeq &\frac{2}{\epsilon}
+ \ln\left(\pi e^{\gamma_{\rm E}} \mu^2 (\theta p)^2\right)
\label{A1num}\\
&-&\frac{11}{72}p^2(\theta p)^2\left(1+\frac{137}{8800}(p^2(\theta p)^2)
+\frac{33}{313600}(p^2(\theta p)^2)^2
+\frac{7129}{17882726400}(p^2(\theta p)^2)^3+...\right)
\nonumber\\
&+&\gamma_{\rm E}\left(1+
\ln \left(\frac{p^2(\theta p)^2}{4}\right)^{\frac{1}{2\gamma_E}}\right)
\frac{p^2(\theta p)^2}{12}
\left(1+\frac{(p^2(\theta p)^2)}{80}+\frac{(p^2(\theta p)^2)^2}{13440}
+\frac{(p^2(\theta p)^2)^3}{3870720}+...\right),
\nonumber
\end{eqnarray}
\begin{eqnarray}
&A_2&=-8\pi^2B
\label{A2num}\\
&\simeq& 2+\frac{7}{18}p^2(\theta p)^2\left(1
+\frac{71}{8400}p^2(\theta p)^2+\frac{1103}{21952000}(p^2(\theta p)^2)^2+
\frac{3587}{19914854400}(p^2(\theta p)^2)^3+...\right)
\nonumber\\
&-&2\gamma_{\rm E}\left(1+
\ln \left(\frac{p^2(\theta p)^2}{4}\right)^{\frac{1}{2\gamma_{\rm E}}}\right)
\frac{p^2(\theta p)^2}{12}\left(1+\frac{p^2(\theta p)^2}{120}
+\frac{(p^2(\theta p)^2)^2}{22400}+\frac{(p^2(\theta p)^2)^3}{6773760}+...\right),
\nonumber
\end{eqnarray}
where $\gamma_{\rm E}\simeq0.577216$ is Euler's constant.
It is to be noted here that the spinor structure proportional to
$\tilde{\fmslash p}$ is missing in the final result, confirming thus
conclusion from \cite{Ettefaghi:2007zz}.

\subsubsection{Divergences and counter terms}

The first striking fact of our closed form result is
the existence of a non-local UV divergence term
\begin{equation}
\Sigma_{\rm UV}=-{\fmslash p}\,\bigg[p^2 \,
\bigg(\frac{\tr\theta\theta}{(\theta p)^2}+2\frac{(\theta\theta p)^2}{(\theta p)^4}\bigg)\bigg]
\cdot\frac{2}{(4\pi)^2\epsilon}\,,
\label{hardUV}
\end{equation}
%containing the additional spinor structure $(\tilde{\tilde p}\equiv\theta\theta p)$.
which does not vanish in the $\theta \to 0$ limit.
This term also  clearly differs with respect to a model not based on Seiberg-Witten maps,
where the UV divergence does not have the momentum and $\theta$ depended factor
$(s_1+2s_2)$. Therefore, the existence of such a divergence suggests a necessity to introduce
the following nonlocal counter-term
\begin{equation}
\Sigma_C=i\delta_2\,\bar\psi\fmslash\partial \,\bigg[\partial^2\,
\bigg(\frac{\tr\theta\theta}{(\theta\partial)^2}
+2\frac{(\theta\theta\partial)^2}{(\theta\partial)^4}\bigg)\bigg]\psi \,,
\label{Ct}
\end{equation}
which would cancel it.

Besides the hard $1/\epsilon$ UV divergence, there is a soft UV/IR mixing term
\cite{Horvat:2011bs}
\begin{equation}
\Sigma_{\rm UV/IR}=-{\fmslash p}\,\bigg[p^2 \,
\bigg(\frac{\tr\theta\theta}{(\theta p)^2}
+2\frac{(\theta\theta p)^2}{(\theta p)^4}\bigg)\bigg]\cdot
\frac{2}{(4\pi)^2}\ln\sqrt{\frac{\mu^2(\theta p)^2}{16}},
\label{lnUV/IR}
\end{equation}
represented by a logarithm,
and it diverges at both the ultraviolet and infrared limits.
Since the soft UV/IR mixing (\ref{lnUV/IR}) appears in (\ref{sigma1A})
with exactly the same coefficient as the UV $1/\epsilon$ term does,
the introduction of the same nonlocal counter-term,
that is (\ref{Ct}), would remove it.
%This certainly appear as new previously unknown
%features of our SW map based $\theta$-exact NCGFT.

Finally, both these terms,
(\ref{hardUV}) and (\ref{lnUV/IR}) respectively, are proportional to $p^2$.
Therefore if the counter term (\ref{Ct}) is included and
the renormalization point is selected at $\fmslash p=0$, our result indicates
that the dispersion relation $p^2=0$ could still hold.
However, in the next subsection we shall investigate the other solutions too.

%%%%%%%%%%%%%%%%%%%%%%%%%%%%%%%%%%%%%%%%%%%%%%%%%%%%%%%%%%%%%%%%%%%%%%%%%%%%%%%%%%%%%%%%%

In the renormalization procedure, all three coefficients
$A_1$, $A_2$ and $B$ from (\ref{sigma1A}) and (\ref{sigma1B})
are renormalized by subtracting counter terms from $\Sigma_1$ in (\ref{sigma1AB}).
We obtain then the renormalized neutrino self energy
\begin{eqnarray}
\Sigma_{1R}&=&-\gamma_{\mu} \bigg[ p^{\mu}\; A_{R} +
(\theta{\theta p})^{\mu}\; \frac{p^2}{(\theta p)^2}\;B_{R}\bigg]\,,
\label{sigma1ABR}\\
A_R&=&\frac{1}{(4\pi)^2} \;\bigg[(s_1+2s_2) A_{1R} + (1+s_1+s_2)A_{2R}\bigg]\,,
\label{1ABR}
\end{eqnarray}
where
\begin{eqnarray}
A_{iR}(p^2,(\theta p)^2,(\theta\theta p)^2)&=&
A_{i}(p^2,(\theta p)^2,(\theta\theta p)^2)-A_{i}(p_0^2,(\theta p)_0^2,(\theta\theta p)_0^2)\,,
\;\;i=1,2;\qquad
\\
B_{R}(p^2,(\theta p)^2,(\theta\theta p)^2)&=&
B(p^2,(\theta p)^2,(\theta\theta p)^2)-B(p_0^2,(\theta p)_0^2,(\theta\theta p)_0^2)
\,,
\label{1ABR0}
\end{eqnarray}
and $(p_0^2,(\theta p)_0^2,(\theta\theta p)_0^2)$ is a choice of renormalization point.

For unitarity \cite{Gomis:2000zz} reasons, it is convenient to take
the following choice of degenerate $\theta$
\begin{equation}
\theta^{\mu\nu}=\frac{1}{\Lambda_{\rm NC}^2}
\begin{pmatrix}
0&0&0&0\\
0&0&1&0\\
0&-1&0&0\\
0&0&0&0
\end{pmatrix}.
\label{degen}
\end{equation}
One finds that $s_1+2s_2=0$ in this degenerate case,
therefore $\Sigma_{\rm UV}=\Sigma_{\rm UV/IR}=0$, so
no counter term is needed. However, note that in this case $(\theta p)^2=0$ for $p_1=p_2=0$,
which renders the modified propagator zero in this subspace.

\subsubsection{Dispersion relation}

In order to probe possible physical consequence of
the one-loop quantum correction $\Sigma_{1-loop}$, we consider the modified propagator
\begin{equation}
\frac{1}{\fmslash\Sigma}=\frac{1}{\fmslash p-\Sigma_{1-loop_M}}=
\frac{\fmslash\Sigma}{\Sigma^2}\,.
\label{disp}
\end{equation}
%and/or the corresponding relativistic Breit-Wigner formula analogy
%\begin{equation}
%\frac{tr\fmslash\Sigma\fmslash\Sigma^*}{|\Sigma^2|^2}=4\frac{|\Sigma|^2}{|\Sigma^2|^2}
%=4\frac{(\re\Sigma)^2 + (\im\Sigma)^2}{(\re\Sigma^2)^2 + (\im\Sigma^2)^2}\,.
%\label{dispBW}
%\end{equation}
The Minkowski counterpart $\Sigma_{1-loop_M}$ of \eqref{Sigma1-loop} is computed
by using the established parameterization technique and Wick rotation.
The resulting one loop correction is
\begin{equation}
\Sigma_{1-loop_M}= e^2\gamma_{\mu}
\bigg[ p^{\mu}\: A + (\theta{\theta p})^{\mu}\; \frac{p^2}{(\theta p)^2}\;B\bigg]\,.
\end{equation}
We further choose the noncommutative parameter to be (\ref{degen})
so that the denominator of \eqref{disp} is finite
and can be expressed explicitly:
\begin{gather}
%(\re\Sigma)^2=p^2\left[\left(\re[\hat A_2]\right)^2\left(\frac{p^4}{p^4_r}
%+2\frac{p^2}{p_r^2}+5\right)-\re[\hat A_2]\left(6+2\frac{p^2}{p_r^2}\right)+1\right]\,,
%\label{dispmod1+}\\
%(\im\Sigma)^2=p^2\left(\Im[\hat A_2]\right)^2\left(\frac{p^4}{p^4_r}
%+2\frac{p^2}{p_r^2}+5\right)\,,
%\label{dispmod1-}\\
\Sigma^2=p^2\left[\hat A_2^2\left(\frac{p^4}{p^4_r}
+2\frac{p^2}{p_r^2}+5\right)-\hat A_2\left(6+2\frac{p^2}{p_r^2}\right)+1\right]\,,
\label{dispmod1}
\end{gather}
where $p_r$ represents $r$-component of the momentum $p$
in a cylindrical spatial coordinate system and $\hat A_2= e^2A_2/(4\pi)^2=-B/2$.

We are interested in the zeros of the denominator,
especially the simple zeros which have the local form $p_0-f(p_i,\theta)$,
for this is associated with the time-evolution of the corresponding
excitation via the Fourier transformation of the propagator
\begin{equation}
\int \prod_i dp_i\int dp_0 \;\frac{\fmslash\Sigma}{\Sigma^2}
\;e^{ip_0({x}-{x'})^0-ip_i( x- x')^i}\,.
\label{Fint}
\end{equation}
In general, a factor $\exp[if(p_i,\theta)t]$ will arise from
through the residue at the zero-point $p_0=f$ which stays in
the upper half of the complex plane.
The existence of the nonzero $\im[f]$ is also natural since it implies
that this excitation decays with respect to time and thus, it is unstable
(the tachyonic modes with $(p_0^2 + m^2)$ type pole can be considered as a special case).
%The real $f$ gives us metastable states with long life time since a slow
%decay factor would be expected when we take $\im \Sigma_2$ corrections into account.

From above one see that $p^2=0$ defines one set of the dispersion relation,
corresponding to the dispersion for the massless neutrino mode,
however the denominator \eqref{dispmod1} has now one more coefficient
\begin{equation}
\Sigma'=\hat A_2^2\left(\frac{p^4}{p^4_r}
+2\frac{p^2}{p_r^2}+5\right)-\hat A_2\left(6+2\frac{p^2}{p_r^2}\right)+1\,,
\label{Sigma'}
\end{equation}
which could also induce certain zero-points. Since the $\hat A_2$ is a
function of a single variable $p^2p_r^2$, with $
p^2=p^2_0-p^2_1-p^2_2-p^2_3\,\,\rm and\,\,p^2_r=p^2_1+p^2_2$,
the condition $\Sigma'=0$ which we are interested in
 can be expressed as a simple algebraic equation
\begin{equation}
\hat A_2^2z^2-2\left(A_2-\hat A_2^2\right)z
+\left(1-6\hat A_2+5\hat A_2^2\right)=0\,,
\label{A2z2}
\end{equation}
of new variables $z:=p^2/p_r^2$, in which the coefficients
are all functions of $y:=p^2p^2_r/\Lambda^4_{\rm NC}$.

The two formal solutions of the equation (\ref{A2z2})
\begin{equation}
z=\frac{1}{\hat A_2}\left[\left(1-\hat A_2\right)\pm\,2\left(\hat A_2
-\hat A_2^2\right)^{\frac{1}{2}}\right]\,,
\label{zsolutions}
\end{equation}
are direction dependent, i.e. birefringent.
The behavior of the birefringent solutions (\ref{zsolutions}),
with respect to a propagating energy, can be analyzed at
two limits $y\to \,0$, and $y\to\,\infty$.\\

%When $y=p^2p^2_r/\Lambda^4_{\rm NC}\ll 1$ we can see that $\im[\hat A_2]\ll\re[\hat A_2]$.
%Also for the worst case if we keep the coupling constant small then
%$\im\Sigma^2\ll\re\Sigma^2$ holds in general. For this reason we can assume that at $y \ll 1$
%\begin{equation}
%i\gamma^{\mu}
%\frac{\re\Sigma_\mu + i\im\Sigma_\mu}{\re\Sigma^2 + i\im\Sigma^2}
%\simeq i\gamma^{\mu}\frac{\re\Sigma_\mu}{\re\Sigma^2}\,,\;\;\;
%\frac{(\re\Sigma)^2 + (\im\Sigma)^2}{(\re\Sigma^2)^2 + (\im\Sigma^2)^2}
%\simeq \frac{(\re\Sigma)^2}{(\re\Sigma^2)^2}\,.
%\label{disp2}
%\end{equation}
%Next we could look at zero points of the denominator, defined by
%$\re\Sigma^2 = 0$. This zeros give us approximately poles of the propagator/the resonance peak.
%Eventually all possible complex zeros gains their meaning by looking at the integral over $p_0$
%in the Fourier transformation

%Here when a pole can be expressed as $p_0-f(p_i,\theta)$,

\noindent
{\it {\bf The low-energy regime: $p^2p^2_r \ll\Lambda^4_{\rm NC}$}}\\
\noindent
For $y \ll 1$ we simply set
$\hat A_2$ and $\hat A_2^2$ to their zeroth order
value $ e^2/8\pi^2$ and $ e^4/64\pi^4$, then
\begin{equation}
z\sim \left(\frac{8\pi^2}{e^2}-1\right)\pm 2\left(\frac{8\pi^2}{e^2}-1\right)^{\frac{1}{2}}\,,
\end{equation}
or equivalently
\begin{equation}
p^2\sim \left(\left(\frac{8\pi^2}{e^2}-1\right)\pm 2
\left(\frac{8\pi^2}{e^2}-1\right)^{\frac{1}{2}}\right)\cdot\,p_r^2
\simeq\left(859\pm 59\right)\cdot\,p_r^2\,,
\label{859pm59}
\end{equation}
defines two (approximate) zero points. From the definition of $p^2$
and $p_r^2$ we see that both solutions are real and positive.
Taking into account the higher order (in y) correction
%from $\im {\hat A}_2$ and $\im {\hat A}_2^2$
these poles will locate nearby the real axis of
the complex $p_0$ plane thus correspond to some metastable modes with
the above defined dispersion relations. As we can see, 
the modified dispersion relation \eqref{859pm59} does not depend on 
the noncommutative scale, therefore it introduces a discontinuity in 
the $\Lambda_{\rm NC}\to\infty$ limit, 
which is not unfamiliar in noncommutative theories.\\

\noindent
{\it {\bf The high-energy regime: $p^2p^2_r \gg \Lambda^4_{\rm NC}$}}\\
At $y\gg 1$ we analyze the asymptotic behavior of $A_2$ starting with its integral form
%\newpage
%$\re[A_2]\sim \im [A_2] \sim \sqrt y$, therefore the above analysis no longer holds,
%one must study the $\Sigma^2$ explicitly in this case.
%So, at $y\gg 1$ limit we do that by replacing $\re[\hat A_2]$
%and $\re[\hat A_2^2]$ by their leading order asymptotic expansion.
%Such expansion could be obtained by recalling the integral form of $A_2$
%(after inverse Wick rotation) at $D=4-\epsilon$
%{\bf
\begin{eqnarray}
A_2&=&\frac{\pi}{2\sin\frac{\epsilon}{2}\pi}
\left(\frac{p_r^2}{\Lambda_{\rm NC}^4}\right)^{\frac{\epsilon}{2}}\int\,dx\,(1-x)
\nonumber\\
&\cdot&
\bigg[\left(-\frac{x(1-x)p^2p_r^2}{\Lambda_{\rm NC}^4}\right)^{1-\frac{\epsilon}{2}}
\Gamma\left(\frac{1}{2}\right){}_1{\tilde F}_2\left(\frac{1}{2};
\frac{3}{2},2-\frac{\epsilon}{2};-\frac{x(1-x)p^2p_r^2}{4\Lambda_{\rm NC}^4}\right)
\nonumber\\
&-&
2^{2-\epsilon}\Gamma\left(-\frac{1}{2}+\frac{\epsilon}{2}\right){}_1{\tilde F}_2
\left(-\frac{1}{2}+\frac{\epsilon}{2};
\frac{\epsilon}{2},\frac{1}{2}+\frac{\epsilon}{2};
-\frac{x(1-x)p^2p_r^2}{4\Lambda_{\rm NC}^4}\right)\bigg]
\nonumber\\
&=&
\left(\frac{p_r^2}{\Lambda_{\rm NC}^4}\right)^\frac{\epsilon}{2}2^{-1-\epsilon}
\frac{\pi^\frac{3}{2}}{\sin\frac{\epsilon}{2}\pi}
\bigg[\left(-\frac{p^2 p_r^2}{16\Lambda_{\rm NC}^4}\right)^{1-\frac{\epsilon}{2}}
\Gamma\left(\frac{1}{2}\right){}_1{\tilde F}_2\left(\frac{1}{2};
\frac{3}{2},\frac{5}{2}-\frac{\epsilon}{2};-\frac{p^2 p_r^2}{16\Lambda_{\rm NC}^4}\right)
\nonumber\\
&-&
\Gamma\left(-\frac{1}{2}
+\frac{\epsilon}{2}\right){}_2{\tilde F}_3\left(-\frac{1}{2}+\frac{\epsilon}{2},1;
\frac{\epsilon}{2},\frac{1}{2}+\frac{\epsilon}{2},
\frac{3}{2};-\frac{p^2 p_r^2}{16\Lambda_{\rm NC}^4}\right)\bigg]\,.
\label{3.41}
\end{eqnarray}
Using the documented asymptotic expansion of generalized
hyper-geometric series \cite{regularizedgeneralizedhypergeometric},
%we have at leading
the leading and next-to-leading asymptotic orders of $A_2$ when $y\to\infty$ reads
\begin{equation}
A_2\sim \frac{i\pi^2}{8}y^{\frac{1}{2}}\left(1-16i\pi^{-1} y^{-1}
e^{-\frac{i}{2}y^\frac{1}{2}}\right)+\mathcal O\left(y^{-1}\right).
\end{equation}
So at leading asymptotic order
\begin{equation}
z\sim -1\pm 2i,
\end{equation}
or
\begin{equation}
p^2_0\sim p^2_3\pm 2i p^2_r.
\end{equation}
We thus reach two unstable deformed modes besides the usual mode
$p^2=0$ in the high energy regime. Here again the leading order
deformed dispersion relation does not depend on the noncommutative scale $\Lambda_{\rm NC}$.

\subsection{Neutrino self-energy for alternative action}

Using the Feynman rule (\ref{V2alt2}) of the alternative action,
we find the following closed form contribution
to the neutrino self-energy from diagram $\Sigma_1$
\begin{equation}
\Sigma_{1_{alt2}}=\frac{\fmslash p}{(4\pi)^2} \bigg[\frac{8}{3} \frac{1}{{(\theta p)^2}}
\bigg(\frac{\tr\theta\theta}{(\theta p)^2}
+4\frac{(\theta\theta p)^2}{(\theta p)^4}\bigg)\bigg]
%\sim {\fmslash p}\;\frac{1}{(|\theta| p^2)^2}
\,.
\label{A9}
\end{equation}
The detailed computation is presented in  Appendix B. From \eqref{V3alt2}
one gets the relation, $g_{\mu\nu}V^{\mu\nu}_{3_{alt2}}(p,-p,q,q)=0$,
showing that $\Sigma_2=0$, while $\Sigma_3$ and $\Sigma_4$
vanish due to charge conjugation symmetry.
Therefore we have again $\Sigma_{1-loop_{alt2}}=\Sigma_{1_{alt2}}$.
There is no alternative dispersion relation in degenerate case (\ref{degen}), since
the factor that multiplies $\fmslash p$ in (\ref{A9}),
does not dependent on the time-like component $p_0$ (energy).
%Using the same technique for the action 1 and/or 3 we found
%the dispersion relation analogy to be $p_r^2=- e^2\Lambda_{\rm NC}^4/3\pi^2$
%which is clearly unsound therefore no new relation is available for alternative action 2.

We have to notice that equation (\ref{A9}) is much simpler than
the corresponding expression (\ref{sigma1AB}).
This is not unfamiliar for actions arising from different SW map deformations.
There are no hard
$1/\epsilon$ UV divergent and no logarithmic UV/IR mixing terms,
and the finite terms like in $A_1$ and $A_2$ are also absent.
Thus the subgraph $\Sigma_1$ does not require any counter-term.
However, the result of the subgraph $\Sigma_1$ evaluation, from
alternative action 2 (\ref{A6}), does express powerful UV/IR mixing effect
due to scale dependent $\theta$-ratios.
%with regard to scale-independent $\theta$-ratios in (\ref{s1s2}).
Namely, in terms of scales only, the $\Sigma_{1_{alt2}}$ experience
the forth-power of the {\it NC-scale/momentum-scale} ratios
$\sim |p|^{-2}|\theta p|^{-2}$ in (\ref{A9}),
i.e. we are dealing with
the $\Sigma_{1_{alt2}}\sim{\fmslash p}\left({\Lambda_{\rm NC}}/{p}\right)^4$
within the ultraviolet and infrared limits for $\Lambda_{\rm NC}$ and $p$, respectively.

The absence of new spinor structure in the alternative neutrino self-energy (\ref{A9})
further suggests the possibility of an appropriate field strength
renormalization with suitable divergence cancellation for $\theta \to 0$ limit.
Here certain hint may be found in the counter terms proposed for
the translation invariant renormalizable noncommutative
$\phi^4$ model with regular commutative limit \cite{Magnen:2008pd}.

\section{Discussion and conclusion}

In this article we present a $\theta$-exact evaluation of the
one-loop quantum correction to the neutral fermion propagator.
We in particular evaluate  the neutrino two-point function in a two different
Seiberg-Witten map based NC$\nu$QED models.
As there is no a priory way to exclude either model, we study them both
and point out their different behavior.
Our method, based on $\theta$-exact expressions of Seiberg-Witten maps,
together with a combination of Schwinger, Feynman, and  HQET parameterization,
yields the one-loop quantum correction in a closed form  for both models.

From the neutrino one-loop two-point function we obtain the self-energy
of a massless neutrino and a dispersion relations which depends on  spacetime
noncommutativity. The general expression for the neutrino self-energy
given in (\ref{sigma1AB}) contains both a hard $1/\epsilon$
ultraviolet term (\ref{hardUV}) and  the celebrated UV/IR mixing term
with a logarithmic infrared singularity $\ln\sqrt{(\theta p)^2}$.
The later reflects the fact that the UV divergence is at most logarithmic, i.e.
there is a soft ultraviolet/infrared term representing UV/IR mixing (\ref{lnUV/IR}),
which is similar to the logarithmic term in the usual vacuum polarization
of the photon, in simple $\star$-product
based NCQED without SW maps \cite{Hayakawa:1999zf}.
Since we have already discussed in detail
properties of the UV/IR mixing term (\ref{lnUV/IR}) in \cite{Horvat:2011bs}
we shall not repeat that discussion here.

The essential difference of our results (\ref{sigma1AB}) as compared to
\cite{Hayakawa:1999zf,Hayakawa:1999yt,Hayakawa:2000zi,Brandt:2001ud,Brandt:2002if}
is that in our case both terms are proportional to the spacetime
noncommutativity dependent $\theta$-ratio factor $(s_1+2s_2)$,
which arise from the natural non-locality of our actions and does not depend on
the noncommutative scale, but only on the scale-independent structure
of the noncommutative $\theta$-ratios.
This behavior, being non-perturbative in nature, differs from that of
fermions in the fundamental or in the adjoint representation in usual,
$\star$-product only based and $\theta$-unexpanded NCQED.
Besides the divergent terms, a new spinor structure $(\theta\theta p)$
with finite coefficients emerges in our computation, see (\ref{sigma1AB})-(\ref{s1s2}).
All these structures are proportional to $p^2$, therefore if appropriate
renormalization conditions are imposed, the commutative dispersion relation
$p^2=0$ can still hold, as a part of the full set of solutions obtained
in (\ref{dispmod1}). Some of the propagating neutrino modes acquire mass.
As the mass depends on the direction of propagation with respect to the
noncommutative background set by $\theta^{\mu\nu}$,
these modes are birefringent, thus confirming
previous result for chiral fermions in NCQED
at first order in $\theta$ \cite{Buric:2010wd}.

The alternative action (\ref{A6}),
presented in section~2.3,  has the same
field content and gauge symmetry as the action (\ref{Sfermion}),
but a different choice of deformation freedom.
%however it is not related by a field redefinition to (\ref{Sfermion}).
Consequently these two different $\theta$-exact actions led to
two different neutrino self-energies (\ref{sigma1AB}) and (\ref{A9}), respectively.
For the unitary choice (\ref{degen}), and in the limit $\theta\to 0$,
self-energy (\ref{sigma1AB}) is finite, however the alternative one, (\ref{A9}),
clearly diverge. This fact indicates strongly that the above two actions
are not related by a field redefinition\footnote{It was shown for NCQED
that only at first order in $\theta$
the Seiberg-Witten map is a field redefinition,
while at higher order in $\theta$ the SW map can not be
regarded as the field redefinition \cite{Grimstrup:2002af}. This is certainly
also true for the SW $\theta$-exact models.
For $\theta$-exact models different SW maps are very much like
different quantization procedures. See discussion of this issue in the Introduction.}.
The corresponding alternative neutrino self-energy (\ref{A9}),
has less striking features than (\ref{sigma1AB}),
but it does have it's own advantages owing to
the absence of a hard UV divergences and the absence of complicated finite terms.
Also, there is no modification of neutrino
dispersion relation from (\ref{A9}) in degenerate case (\ref{degen}).
The structure in (\ref{A9}) is different
(it is {\it NC-scale/energy} dependent) with respect to
the scale-independent structure $(s_1+2s_2)$ from (\ref{sigma1A12}),
as well as to the structure arising from fermion self-energy computation in
the case of $\star$-product only unexpanded theories \cite{Hayakawa:1999zf}.
However, (\ref{A9}) does posses powerful UV/IR mixing effect.
This is fortunate with regard to the use of  low-energy NCQFT
as an important window  to quantum gravity \cite{Szabo:2009tn}
and holography \cite{Horvat:2010km}.

The low energy dispersion relation \eqref{859pm59} is, 
in principle, capable of generating a direction dependent superluminal velocity, 
this can be seen clearly from the maximal attainable velocity of the neutrinos
\begin{equation}
\frac{{\rm v}_{max}}{c}=\frac{dE}{d|\vec p|}\sim \sqrt{
1+\left(859\pm 59\right)
\sin^2\vartheta}\,,
\label{varepsilon}
\end{equation}
where $\vartheta$ is the angle with respect to the direction perpendicular 
to the NC plane. This gives one more example how such spontaneous breaking 
of Lorentz symmetry (via the $\theta$-background) could affect 
the particle kinematics through quantum corrections 
(even without divergent behavior like UV/IR mixing). 
On the other hand one can also see that the magnitude of superluminosity 
is in general very large in our model, thus seems contradicting various observations.
\footnote{The currently largest superluminal velocity for neutrinos, 
$({\rm v} -c)/c = (2.37 \pm 0.32 ({\rm stat.}){+0.34 \atop -0.24} ({\rm sys.})) 
\times 10^{-5}$ reported by the OPERA collaboration \cite{:2011zb}, 
has been found to be suffering from multiple technical difficulties lately 
\cite{OPERAnews}. Other experiments suggests much smaller values 
\cite{Hirata:1987hu,Bionta:1987qt,Longo:1987ub}.} 
The authors consider here, on the other hand, 
that the large superluminal velocity issue may be reduced/removed by 
taking into account several further considerations:

\begin{itemize}
\item The reason to select a constant nonzero $\theta$ 
background in this paper is computational simplicity. 
The results will, however, still hold for a NC background 
that is varying sufficiently slowly with respect to the scale of
noncommutativity. There is no physics reason to expect 
$\theta$ to be a globally constant background {\it ether}.  
In fact, if the $\theta$ background is only nonzero in 
tiny regions (NC bubbles) the effects of the modified dispersion
relation will be suppressed macroscopically.
Certainly a better understanding of possible sources of NC is needed.

\item In our computation we considered only the purely noncommutative 
neutrino-photon coupling, it has been pointed out that modified 
neutrino dispersion relation could open decay channels within 
the commutative standard model framework \cite{Cohen:2011hx}. 
In our case this would further provide decay channel(s) which can 
bring superluminal neutrinos to normal ones.

\item Finally, as we have stated in the introduction, 
model 1 is not the only allowed deformed model with noncommutative 
neutrino-photon coupling. And as we have shown for our model 2, 
there could be no modified dispersion relation(s) for deformation(s) 
other than 1, therefore it is reasonable to conjecture 
that Seiberg-Witten map freedom may also serve as one possible 
remedy to this issue.
\end{itemize}

\section{Acknowledgment}
J.T. would like to acknowledge support of Alexander von Humboldt Foundation
(KRO 1028995), Max-Planck-Institute for Physics, Munich, for hospitality,
and W. Hollik for fruitful discussions.
The work of R.H. and J.T. are supported by
the Croatian Ministry of Science, Education and Sports
under Contracts Nos. 0098-0982930-2872 and 0098-0982930-2900, respectively.
The work of A.I. is supported by the Croatian Ministry of Science, Education and Sports
under Contracts Nos. 0098-0982930-1016.
The work of J.Y. was supported by the NSF and IRB Zagreb, Croatia.
%and by the German Research Foundation (Deutsch
%Forschungsgemeinschaft (DFG)) through the Institutional Strategy of the
%University of G\"ottingen.

%\section{One point function and charge conjugation symmetry}\label{discrete}
\appendix
\section{Loop integrals}
\subsection{Integral $I_{123}$}
Integral $I_{123}$ follow the same computation in NCQED without SW map \cite{Hayakawa:1999zf},
resulting in
\begin{eqnarray}
I_{123}&=&\frac{1}{(4\pi)^{\frac{D}{2}}}2\fmslash p\bigg\{(p^2)^{\frac{D}{2}-2}
\Gamma\left(2-\frac{D}{2}\right)B(\frac{D}{2}-1,\frac{D}{2})
\label{123'}\\
&-&2\int\limits_0^1 dx(1-x)\left(x(1-x)p^2\right)^{\frac{D}{4}-1}
\cdot 2^{\frac{D}{2}-2}(\tilde p^2)^{1-\frac{D}{4}}K_{2-\frac{D}{2}}
\left[\left(x(1-x)p^2\tilde
p^2\right)^{\frac{1}{2}}\right]\bigg\}\,.
\nonumber
\end{eqnarray}

\subsection{Integral parameterizations for $I_{456}$ and $I_{789}$}
Next part involves integrals $I_{456}$ and $I_{789}$
which differ from $I_{123}$ by the existence of a non-quadratic
$q\theta p$ denominators. To overcome this problem we introduce
the HQET parameterization \cite{Grozin:2000cm}, represented as follows
\begin{equation}
\frac{1}{a_1^{n_1} a_2^{n_2}}=
 \frac{\Gamma(n_1+n_2)}{\Gamma(n_1)\Gamma(n_2)}
 \int_0^\infty\frac{i^{n_1}y^{n_1-1} dy}{(ia_1y + a_2)^{n_1+n_2}}\,.
\label{Grozin}
\end{equation}

To perform computations of integrals
(\ref{123}), (\ref{456}) and (\ref{789}), we first use
the Feynman parameterization on the quadratic denominators,
then the HQET parameterization help us to combine
the quadratic and linear denominators.
So for the denominators in $I_{456}$ we have
\begin{eqnarray}
%\begin{split}
\frac{1}{q^2(p+q)^2}\frac{1}{q\theta p}&=&2i\int\limits_0^1
dx\int\limits_0^\infty dy
\frac{1}{\left[(q^2+i\epsilon)(1-x)+((p+q)^2+i\epsilon)x+iy(q\theta
p)\right]^3}
\nonumber\\
&=&2i\int\limits_0^1 dx\int\limits_0^\infty dy
\frac{1}{\left[(q+xp+\frac{i}{2}y\tilde
p)^2+x(1-x)p^2+\frac{y^2}{4}\tilde p^2+i\epsilon\right]^3}\,.
%\end{split}
\label{denom456}
\end{eqnarray}

Now we use the Schwinger parameterization to turn the denominators
into Gaussian integrals. This then combines with different phase factors.
For the zero phase, after an integral over the loop-momenta
$l=q+xp+\frac{i}{2}y\tilde p$ and
changing variable $y$ to $Y=\alpha y$ we have
\begin{eqnarray}
%\begin{split}
I_4&=&2i\int\limits_0^1 dx\int\limits_0^\infty dy\int \frac{d^D
l}{(2\pi)^D}\frac{\left[l^2\left(\frac{2}{D}-2\right)/\!\!\!\tilde
p-iy(1-x)/\!\!\,\tilde\tilde \!\!p \,p^2+\frac{y^2}{2}/\!\!\!\tilde
p \tilde p^2\right]}{\left[l^2+x(1-x)p^2+\frac{y^2}{4}\tilde
p^2+i\epsilon\right]^3}
\nonumber\\
&=&i\int\limits_0^1 dx\int\limits_0^\infty dY\int \frac{d^D
l}{(2\pi)^D}\int\limits_0^\infty d\alpha
\alpha\left[l^2\left(\frac{2}{D}-2\right)/\!\!\!\tilde
p-\frac{i}{\alpha}Y(1-x)/\!\!\,\tilde\tilde \!\!p\,
p^2+\frac{Y^2}{2\alpha^2}/\!\!\!\tilde p \tilde p^2\right]
\nonumber\\
&\cdot&\exp\left[-\alpha l^2-\alpha
x(1-x)p^2-\frac{Y^2}{4\alpha}\tilde p^2+i\epsilon\right]\,.
%\end{split}
\label{I4}
\end{eqnarray}
For the other two nonzero phase factors,
one can first performe the Schwinger parameterization
like in the zero-phase case, then change
variable $y$ to $Y=\alpha\left(y\pm\frac{1}{\alpha}\right)$.
The resulting integrals are
\begin{eqnarray}
%\begin{split}
I_{5.5\pm0.5}&=&i\int\limits_0^1 dx\int\limits_{\pm 1}^\infty dY\int \frac{d^D
l}{(2\pi)^D}\int\limits_0^\infty d\alpha
\alpha\left[l^2\left(\frac{2}{D}-2\right)/\!\!\!\tilde
p-\frac{i}{\alpha}Y(1-x)/\!\!\,\tilde\tilde \!\!p\,
p^2+\frac{Y^2}{2\alpha^2}/\!\!\!\tilde p \tilde p^2\right]
\nonumber\\
&\cdot&\exp\left[-\alpha l^2-\alpha
x(1-x)p^2-\frac{Y^2}{4\alpha}\tilde p^2+i\epsilon\right].
%\end{split}
\label{5.5}
\end{eqnarray}
One can easily see that they differ from the zero-phase case just by a different
$Y$ integral limit. Summing over these three terms
and integrating over the loop-momenta, one obtains
\begin{eqnarray}
%\begin{split}
I_{456}&=&2I_4-(I_5+I_6)
\nonumber\\
&=&2\frac{1}{(4\pi)^{\frac{D}{2}}}\int\limits_0^1
dx\int\limits_0^1 dY\int\limits_0^\infty d\alpha
\alpha^{-\frac{D}{2}}Y(1-x)/\!\!\,\tilde\tilde \!\!p\,
p^2\cdot\exp\left[-\alpha x(1-x)p^2-\frac{Y^2}{4\alpha}\tilde p^2\right]
%\\
%=&\frac{1}{(4\pi)^{\frac{D}{2}}}\int\limits_0^1 dx\int\limits_0^1
%dY^2\int\limits_0^\infty d\alpha
%\alpha^{-\frac{D}{2}}(1-x)/\!\!\!\,\tilde\tilde \!\!p
%p^2\cdot\exp\left[-\alpha x(1-x)p^2-\frac{Y^2}{4\alpha}\tilde
%p^2\right]
\nonumber\\
&=&\frac{1}{(4\pi)^{\frac{D}{2}}}\frac{4/\!\!\,\tilde\tilde \!\!p\,
p^2}{\tilde p^2}\bigg\{\int\limits_0^1 dx\int\limits_0^\infty
d\alpha\alpha^{1-\frac{D}{2}}(1-x)\exp\left[-\alpha x(1-x)p^2\right]
\nonumber\\
&-&\int\limits_0^1 dx\int\limits_0^\infty
d\alpha\alpha^{1-\frac{D}{2}}(1-x)\exp\left[-\alpha
x(1-x)p^2-\frac{\tilde p^2}{4\alpha}\right]\bigg\}.
%\end{split}
\label{I456}
\end{eqnarray}
The integral over $\alpha$ is then straightforward, so we have
\begin{eqnarray}
%\hspace{-0.25in}
%\begin{split}
I_{456}&=&\frac{1}{(4\pi)^{\frac{D}{2}}}\frac{4\;/\!\!\,\tilde\tilde
\!\!p\, p^2}{\tilde p^2}\bigg\{\int\limits_0^1 dx\int\limits_0^\infty
d\alpha\alpha^{1-\frac{D}{2}}(1-x)\exp\left[-\alpha x(1-x)p^2\right]
\nonumber\\
&-&\int\limits_0^1 dx\int\limits_0^\infty
d\alpha\alpha^{1-\frac{D}{2}}(1-x)\exp\left[-\alpha
x(1-x)p^2-\frac{\tilde p^2}{4\alpha}\right]\bigg\}
\label{I456-}\\
&=&\frac{1}{(4\pi)^{\frac{D}{2}}}\frac{4\;/\!\!\,\tilde\tilde \!\!p\,
p^2}{\tilde
p^2}\bigg\{(p^2)^{\frac{D}{2}-2}\Gamma\left(2-\frac{D}{2}\right)B
\left(\frac{D}{2}-1,\frac{D}{2}\right)
\nonumber\\
&-&2\int\limits_0^1 dx(1-x)\left(x(1-x)p^2\right)^{\frac{D}{4}-1}
\cdot 2^{\frac{D}{2}-2}(\tilde p^2)^{1-\frac{D}{4}}K_{2-\frac{D}{2}}
\left[\left(x(1-x)p^2\tilde
p^2\right)^{\frac{1}{2}}\right]\bigg\}.
%\end{split}
\nonumber
\end{eqnarray}
%\subsection{$I_{7,8,9}$}
%Here is the computation for
%first in Euclidian and $(x, iy, \alpha)$ parameterization.
%Under this parameterization, we have

\subsection{Integral $I_{789}$}

%Our next task is to evaluate the integral $I_{789}$.
A parameterization procedure introduced in the previous subsection yields
the following outcome for the integral $I_{789}$:
\begin{eqnarray}
%\begin{split}
I_{789}&=&-2\int\frac{d^Dl}{(2\pi)^D}\int\limits_0^1
dx\int\limits_0^{\frac{1}{\alpha}} dY\int\limits_0^\infty d\alpha
\alpha^3\exp\left[-\alpha\left[\left(l^2+i\epsilon\right)+x(1-x)p^2+Y^2\frac{\tilde
p^2}{4}\right]\right]
\nonumber\\
&\cdot&\bigg\{\bigg[l^2(1-x)Y\left(\tilde p^2\fmslash
p\left(\frac{2}{D}-1\right)+p^2/\!\!\!\tilde{\tilde
p}\frac{2}{D}-\frac{\tr\theta\theta}{D}p^2\fmslash p\right)
\nonumber\\
&-&Y^3(1-x)\frac{\fmslash p}{4}\left(\tilde{\tilde p}^2p^2-\tilde
p^4\right)\bigg]-\frac{1}{\alpha}\bigg[l^2(1-x)\left(\tilde
p^2\fmslash p\left(\frac{2}{D}-1\right)+p^2/\!\!\!\tilde{\tilde
p}\frac{2}{D}-\frac{\tr\theta\theta}{D}p^2\fmslash p\right)
\nonumber\\
&-&Y^2(1-x)\frac{\fmslash p}{4}\left(\tilde{\tilde p}^2p^2-\tilde
p^4\right)\bigg]\bigg\}\,.
%\end{split}
\label{I789}
\end{eqnarray}
The rest of the integrals are much more complicated but can still be handled.
The integral $I_{789}$ after the Gaussian integration over loop-momenta $l$ reads as follows:
\begin{eqnarray}
%\begin{split}
&&I_{789}=-\frac{2}{(4\pi)^{\frac{D}{2}}}\int\limits_0^1
dx\int\limits_0^{\frac{1}{\alpha}} dY\int\limits_0^\infty d\alpha
\exp\left[-\alpha\left(x(1-x)p^2+Y^2\frac{\tilde
p^2}{4}\right)\right]
\label{B1}\\
&&\cdot\bigg\{\bigg[(1-x)Y\alpha^{2-\frac{D}{2}}\left(\tilde p^2\fmslash
p\left(1-\frac{D}{2}\right)+p^2/\!\!\!\tilde{\tilde
p}-\frac{\tr\theta\theta}{2}p^2\fmslash p\right)
%\\&
-(1-x)Y^3\alpha^{3-\frac{D}{2}}\frac{\fmslash
p}{4}\left(\tilde{\tilde p}^2p^2-\tilde
p^4\right)\bigg]
\nonumber\\
&&-\bigg[(1-x)\alpha^{1-\frac{D}{2}}\left(\tilde
p^2\fmslash p\left(1-\frac{D}{2}\right)+p^2/\!\!\!\tilde{\tilde
p}-\frac{\tr\theta\theta}{2}p^2\fmslash p\right)
%\\&
-(1-x)Y^2\alpha^{2-\frac{D}{2}}\frac{\fmslash
p}{4}\left(\tilde{\tilde p}^2p^2-\tilde p^4\right)\bigg]\bigg\}\,.
%\end{split}
\nonumber
\end{eqnarray}
To evaluate (\ref{B1}), one first has to separate it into two parts
\begin{equation}
I_{789}=\frac{1}{(4\pi)^{\frac{D}{2}}}
\frac{\mathcal{I}+\mathcal{I}'}{2}\,,
\label{B2}
\end{equation}
where
\begin{eqnarray}
%\begin{split}
\mathcal{I}&=&-2\int\limits_0^1 dx\int\limits_0^{\frac{1}{\alpha}}
dY\int\limits_0^\infty d\alpha
\exp\left[-\alpha\left(x(1-x)p^2+Y^2\frac{\tilde
p^2}{4}\right)\right]
\label{B3}\\
&\cdot&\bigg[(1-x)Y\alpha^{2-\frac{D}{2}}\left(\tilde p^2\fmslash
p\left(1-\frac{D}{2}\right)+p^2/\!\!\!\tilde{\tilde
p}-\frac{\tr\theta\theta}{2}p^2\fmslash
p\right)
%\\&
-(1-x)Y^3\alpha^{3-\frac{D}{2}}\frac{\fmslash
p}{4}\left(\tilde{\tilde p}^2p^2-\tilde p^4\right)\bigg]\,,
%\end{split}
\nonumber\\
%\end{equation}
%and
%\begin{equation}
%\begin{split}
%\mathcal{I}'=&2\int\limits_0^1 dx\int\limits_0^{\frac{1}{\alpha}}
%dY\int\limits_0^\infty d\alpha
%\exp\left[-\alpha\left(x(1-x)p^2+Y^2\frac{\tilde
%p^2}{4}\right)\right]
%\\
%&\cdot\bigg[(1-x)\alpha^{1-\frac{D}{2}}\left(\tilde p^2\fmslash
%p\left(1-\frac{D}{2}\right)+p^2/\!\!\!\tilde{\tilde
%p}-\frac{\tr\theta\theta}{2}p^2\fmslash
%p\right)
%%\\&
%-(1-x)Y^2\alpha^{2-\frac{D}{2}}\frac{\fmslash
%p}{4}\left(\tilde{\tilde p}^2p^2-\tilde p^4\right)\bigg]\,.
%\end{split}
\mathcal{I}'&=& \mathcal{I}\left(Y\alpha^{2-\frac{D}{2}}
\to \alpha^{1-\frac{D}{2}}\right)\,.
\label{B4}
\end{eqnarray}
The integral $\mathcal I$ can be evaluated in the same way as $I_{456}$ since
\begin{eqnarray}
%\begin{split}
\mathcal{I}&=&-\int\limits_0^1 dx\int\limits_0^{\frac{1}{\alpha^2}}
dY^2\int\limits_0^\infty d\alpha
\exp\left[-\alpha\left(x(1-x)p^2+Y^2\frac{\tilde
p^2}{4}\right)\right]
\label{B5}\\
&\cdot&\bigg[(1-x)\alpha^{2-\frac{D}{2}}\left(\tilde p^2\fmslash
p\left(1-\frac{D}{2}\right)+p^2/\!\!\!\tilde{\tilde
p}-\frac{\tr\theta\theta}{2}p^2\fmslash
p\right)
%\\&
-(1-x)Y^2\alpha^{3-\frac{D}{2}}\frac{\fmslash
p}{4}\left(\tilde{\tilde p}^2p^2-\tilde p^4\right)\bigg]\,.
%\end{split}
\nonumber
\end{eqnarray}
The resulting $Y$ and $\alpha$ integration are given in terms of the
integrals over the modified Bessel functions $K_{2-\frac{D}{2}}[x]$ and
$K_{1-\frac{D}{2}}[x]$, respectively:
\begin{eqnarray}
%\begin{split}
&&\mathcal I=-4\bigg[\left(\fmslash
p\left(1-\frac{D}{2}\right)+\frac{p^2/\!\!\!\tilde{\tilde p}}{\tilde
p^2}-\frac{\tr\theta\theta}{2}\frac{p^2\fmslash p}{\tilde
p^2}\right)-\frac{\fmslash p}{\tilde p^4}\left(\tilde{\tilde
p}^2p^2-\tilde
p^4\right)\bigg]
\label{B6}\\
&&\cdot\bigg\{(p^2)^{\frac{D}{2}-2}\Gamma\left(2-\frac{D}{2}\right)B
\left(\frac{D}{2}-1,\frac{D}{2}\right)
\nonumber\\
&&-2\int\limits_0^1 dx(1-x)\left(x(1-x)p^2\right)^{\frac{D}{4}-1}
\,2^{\frac{D}{2}-2}(\tilde p^2)^{1-\frac{D}{4}}K_{2-\frac{D}{2}}
\left[\left(x(1-x)p^2\tilde
p^2\right)^{\frac{1}{2}}\right]\bigg\}
\nonumber\\
&&-2\frac{\fmslash p}{\tilde p^2}\left(\tilde{\tilde
p}^2p^2-\tilde p^4\right)
%\\&\cdot
\int\limits_0^1 dx(1-x)
\left(x(1-x)p^2\right)^{\frac{D}{4}-\frac{1}{2}}
\,2^{\frac{D}{2}-1}(\tilde p^2)^{\frac{1}{2}
-\frac{D}{4}}K_{1-\frac{D}{2}}\left[\left(x(1-x)p^2\tilde
p^2\right)^{\frac{1}{2}}\right]\,.
%\end{split}
\nonumber
\end{eqnarray}
Next we evaluate the integral $\mathcal{I}'$ from (\ref{B4}),
where one could again first integrate over $Y$, which yields both, the error function
$\rm{erf}[x]$ and the modified Bessel function $K_{1-\frac{D}{2}}[x]$, respectively
\begin{eqnarray}
&&\mathcal{I}'=\pi^{\frac{1}{2}}
\bigg\{2\left(\tilde p^2\fmslash
p\left(1-\frac{D}{2}\right)+p^2/\!\!\!\tilde{\tilde
p}-\frac{\tr\theta\theta}{2}p^2\fmslash
p\right)-\frac{\fmslash p}{\tilde p^2}
\left(\tilde{\tilde p}^2p^2-\tilde p^4\right)\bigg\}
\label{B8}\\
&&\cdot(\tilde p^2)^{-\frac{1}{2}}\int\limits_0^1 dx(1-x)\int\limits_0^\infty
d\alpha\alpha^{\frac{1}{2}-\frac{D}{2}}
\cdot\exp\left[-\alpha\left(x(1-x)p^2\right)\right]\, \rm{erf}
\left[\frac{(\tilde p^2)^\frac{1}{2}}{2\alpha^\frac{1}{2}}\right]
\nonumber\\
&&+2\frac{\fmslash p}{\tilde p^2}\left(\tilde{\tilde p}^2p^2-\tilde p^4\right)
%\\&\cdot
\int\limits_0^1 dx(1-x)\left(x(1-x)p^2\right)^{\frac{D}{4}-\frac{1}{2}}
\, 2^{\frac{D}{2}-1}(\tilde p^2)^{\frac{1}{2}-\frac{D}{4}}K_{1-\frac{D}{2}}
\left[\left(x(1-x)p^2\tilde
p^2\right)^{\frac{1}{2}}\right]\,.
\nonumber
\end{eqnarray}
The integral over the error function can be expressed in terms of
a regularized hyper-geometric functions
\begin{eqnarray}
%\begin{split}
&&\pi^{\frac{1}{2}}\int\limits_0^1 dx(1-x)\int\limits_0^\infty
d\alpha\alpha^{\frac{1}{2}-\frac{D}{2}}\cdot\exp\left[-\alpha\left(x(1-x)p^2\right)\right]
\cdot \rm{erf}\left[\frac{(\tilde p^2)^\frac{1}{2}}{2\alpha^\frac{1}{2}}\right]
\nonumber\\
&&=\frac{\pi}{2\sin\frac{D\pi}{2}}
\int\limits_0^1 dx(1-x)(\tilde p^2)^{\frac{1}{2}}
%\nonumber\\
%&&
\cdot\bigg[\left(x(1-x)p^2\right)^{\frac{D}{2}-1}
\Gamma\left(\frac{1}{2}\right)\left(\frac{1}{2};
\frac{3}{2},\frac{D}{2};\frac{x(1-x)p^2\tilde p^2}{4}\right)
\nonumber\\
&&-2^{D-2}(\tilde p^2)^{1-\frac{D}{2}}\Gamma\left(\frac{3-D}{2}\right)
\left(\frac{3-D}{2};\frac{4-D}{2},\frac{5-D}{2};
\frac{x(1-x)p^2\tilde p^2}{4}\right)\bigg]\,.
%\end{split}
\label{B9}
\end{eqnarray}
Therefore, in the end we have
\begin{eqnarray}
%\begin{split}
I_{789}&=&\frac{1}{(4\pi)^{\frac{D}{2}}}\bigg\{-4\bigg[\left(\fmslash
p\left(1-\frac{D}{2}\right)+\frac{p^2/\!\!\!\tilde{\tilde p}}{\tilde
p^2}-\frac{\tr\theta\theta}{2}\frac{p^2\fmslash p}{\tilde
p^2}\right)-\frac{\fmslash p}{\tilde p^4}\left(\tilde{\tilde
p}^2p^2-\tilde
p^4\right)\bigg]
\nonumber\\
&\cdot&\bigg[(p^2)^{\frac{D}{2}-2}\Gamma\left(2-\frac{D}{2}\right)B
\left(\frac{D}{2}-1,\frac{D}{2}\right)
\nonumber\\
&-&2\int\limits_0^1 dx(1-x)\left(x(1-x)p^2\right)^{\frac{D}{4}-1}
2^{\frac{D}{2}-2}(\tilde p^2)^{1-\frac{D}{4}}K_{2-\frac{D}{2}}
\left[\left(x(1-x)p^2\tilde
p^2\right)^{\frac{1}{2}}\right]\bigg]\bigg\}
\nonumber\\
&+&\frac{1}{(4\pi)^{\frac{D}{2}}}\bigg\{2\left(\fmslash
p\left(1-\frac{D}{2}\right)+\frac{p^2/\!\!\!\tilde{\tilde
p}}{\tilde p^2}-\frac{\tr\theta\theta}{2}\frac{p^2\fmslash
p}{\tilde p^2}\right)-\frac{\fmslash
p}{\tilde p^4}\left(\tilde{\tilde p}^2p^2-\tilde p^4\right)\bigg\}
\nonumber\\
&\cdot&\frac{\pi}{2 \sin\frac{D\pi}{2}}
\int\limits_0^1 dx(1-x)(\tilde p^2)^{2-\frac{D}{2}}
\nonumber\\
&\cdot&\bigg[\left(x(1-x)p^2\tilde p^2\right)^{\frac{D}{2}-1}
\Gamma\left(\frac{1}{2}\right){}_1{\tilde F}_2\left(\frac{1}{2};
\frac{3}{2},\frac{D}{2};\frac{x(1-x)p^2\tilde p^2}{4}\right)
\nonumber\\
&-&2^{D-2}\Gamma\left(\frac{3-D}{2}\right){}_1{\tilde F}_2
\left(\frac{3-D}{2};\frac{4-D}{2},\frac{5-D}{2};
\frac{x(1-x)p^2\tilde p^2}{4}\right)\bigg]\,.
%\end{split}
\label{B10}
\end{eqnarray}

\section{Two-point function of alternative model}

In this appendix we compute the one-loop integral $\Sigma_1$
of the alternative model 2, which starts with
\begin{equation}
\Sigma_1=-\fmslash p\int\,\frac{d^Dq}{(4\pi)^D}\frac{\tilde q^2}{q^2}
\frac{\sin^2\left(\frac{q\theta p}{2}\right)}{\left(\frac{q\theta p}{2}\right)^2}
=-\fmslash p\cdot \Sigma\,.
\end{equation}
By the parameterization we split $\Sigma$ into the two integrals, $I$ and $I'$,
respectively:
\begin{equation}
\begin{split}
\Sigma&=-\int\,\frac{d^Dq}{(4\pi)^D}\int\limits_0^\infty\,dy y\tilde q^2\left(2-e^{iq\theta p}
-e^{-iq\theta p}\right)\int\limits_0^\infty\,d\alpha \alpha^2 e^{-\alpha(q^2+iy(q\theta p))}
\\
&=I-I'\,,
\end{split}
\end{equation}
\begin{equation}
I=2\int\,\frac{d^Dl}{(4\pi)^D}\int\limits_0^1\,dY\cdot
Y\int\limits_0^\infty\,d\alpha\left(\tr\theta\theta\cdot\frac{l^2}{D}
+\frac{Y^2\tilde{\tilde p}^2}{4\alpha^2}\right)e^{-\alpha l^2-\frac{Y^2\tilde p^2}{4\alpha}}\,.
\end{equation}
The Gaussian integration over $l$ than produces
\begin{equation}
I=\frac{2}{(4\pi)^{\frac{D}{2}}}\int\limits_0^1\,dY\cdot Y\int\limits_0^\infty\,d\alpha
\left(\frac{\tr\theta\theta}{2}\alpha^{-1-\frac{D}{2}}
+\frac{Y^2\tilde{\tilde p}^2}{4}\alpha^{-2-\frac{D}{2}}\right)e^{-\frac{Y^2\tilde p^2}{4\alpha}}\,,
\end{equation}
\begin{equation}
I'=I(dY\cdot Y \to dY)\,.
\end{equation}
%\begin{equation}
%I'=\frac{2}{(4\pi)^{\frac{D}{2}}}\int\limits_0^1\,dY\int\limits_0^\infty\,d\alpha
%\left(\frac{\tr\theta\theta}{2}\alpha^{-1-\frac{D}{2}}+\frac{Y^2\tilde{\tilde p}^2}{4}
%\alpha^{-2-\frac{D}{2}}\right)e^{-\frac{Y^2\tilde p^2}{4\alpha}}
%\end{equation}
The change of coordinate $\alpha\to\rho=1/\alpha$ and then the integral over $\rho$ gives
\begin{equation}
I=\frac{2}{(4\pi)^{\frac{D}{2}}}
\left[\frac{\tr\theta\theta}{2}\left(\frac{\tilde p^2}{4}\right)^{-\frac{D}{2}}\Gamma
\left(\frac{D}{2}\right)
+\frac{\tilde{\tilde p}^2}{4}
\left(\frac{\tilde p^2}{4}\right)^{-1-\frac{D}{2}}\Gamma\left(1+\frac{D}{2}\right)\right]
\int\limits_0^1\,dY\cdot Y^{1-D}\,,
\end{equation}
\begin{equation}
I'=I(dY\cdot Y^{1-D} \to dY\cdot Y^{-D})\,.
\end{equation}
Now we assume
\begin{equation}
\int\limits_0^1\,dY\cdot Y^{1-D}\equiv\frac{1}{2-D}\,,\;\;
\int\limits_0^1\,dY\cdot Y^{-D}\equiv\frac{1}{1-D}.
\end{equation}
This is clearly incorrect for $D=4$ as the integral diverges at the lower limit.
Here our motivation is from the general practice of dimensional regularization,
in which the integrals could be evaluated at a certain $D$ value where it converges,
then take the extended $D$ value. We then set $D=4$ and finally obtain
\begin{equation}
I=\frac{-1}{(4\pi)^2}\left(8\frac{\tr\theta\theta}{(\theta p)^4}
+32\frac{(\theta\theta p)^2}{(\theta p)^6}\right),\;\;\;
I'= \frac{2}{3}I\,.
\end{equation}

%\bibliography{thesis}
%\bibliographystyle{JHEP}

\end{document}